# Phononic Switching of Magnetization by the Ultrafast Barnett Effect

C. S. Davies[1,2*], F. G. N. Fennema[1,2], A. Tsukamoto[3], I. Razdolski[1,2,4], A. V. Kimel[2] and A. Kirilyuk[1,2]

[1]*FELIX Laboratory, Radboud University, Toernooiveld 7, 6525 ED Nijmegen, The Netherlands*
[2]*Radboud University, Institute for Molecules and Materials, Heyendaalseweg 135, 6525 AJ Nijmegen, The Netherlands*
[3]*College of Science and Technology, Nihon University, 7-24-1 Funabashi, Chiba 274-8501, Japan*
[4]*Faculty of Physics, University of Bialystok, Ciolkowskiego 1L, 15-245 Bialystok, Poland*

*Correspondence to: carl.davies@ru.nl (C.S.D)

**Abstract:** The Barnett effect[1,2], discovered more than a century ago, describes how an inertial body with otherwise zero net magnetic moment acquires spontaneous magnetization when mechanically spinning. Breakthrough experiments have recently shown that an ultrashort laser pulse destroys the magnetization of an ordered ferromagnet within hundreds of femtoseconds[3], with the spins losing angular momentum to circularly-polarized optical phonons as part of the ultrafast Einstein-de Haas effect[4,5]. However, the prospect of using such high-frequency vibrations of the lattice to reciprocally switch magnetization in a nearby magnetic medium has not yet been experimentally explored. Here we show that the spontaneous magnetization temporarily gained via the ultrafast Barnett effect, through the resonant excitation of circularly-polarized optical phonons in paramagnetic substrates, can be used to permanently reverse the magnetic state of the substrate-mounted heterostructure. With the handedness of the phonons steering the direction of magnetic switching, the ultrafast Barnett effect offers a selective and potentially universal method for exercising ultrafast non-local control over magnetic order.

**Main Text:** All condensed phases of matter have a characteristic set of inter-atomic bonds that are capable of stretching, bending, rocking, wagging etc. These microscopic motions of the lattice, from which the entire spectrum of optical phonons derive, are often intimately linked to macroscopic order parameters such as charge, magnetism, polarization, crystallography and superconductivity. By driving specific high-frequency phonons at resonance[6], therefore, one can displacively manipulate coupled order parameters characteristic of the excited system, as demonstrated by experiments in the thriving research field of non-linear phononics[7-8].

The absorption of infrared (IR) light is mediated by IR-active optical phonons. Usually, the collective and coherent oscillations of the dipole moment (a normal mode) along a certain axis (the normal coordinate) is taken to mean that such phonons are linearly-polarized. Pioneering experiments, however, have proven that one can also excite phonons with circular polarization. In two-dimensional dichalcogenides, the symmetry provided by honeycomb lattices allows one to distinguish rotational ionic vibrations that are intrinsically chiral[9]. Alternatively, one can generate left- or right-handed circularly-polarized optical phonons by simultaneously exciting two normal modes out of phase with each other[10]. Such phonons have very recently been identified as absorbers of angular momentum during the nascent stage of ultrafast photo-induced demagnetization[4,5].

The possibility of exciting circularly-polarized phonons leads naturally to the question of whether one can harness these to interface with magnetization[11]. Elliptically-polarized optical phonons in ErFeO$_3$ were inferred as manufacturing an effective magnetic field, acting locally within the antiferromagnet, with amplitude on the order of tens of milli-Tesla[10]. Several theoretical works have since predicted[13-15] that circularly-polarized optical phonons, driven at resonance, can generate so-called "phono-magnetic" fields in a wide variety of materials with different point group symmetries and magnetic properties. By virtue of the relatively long lifetime and non-linear character of optical phonons, combined with the possibility of driving the circularly-polarized phonons with a multi-cycle excitation at resonance, the phono-magnetic field has been predicted to become as strong as several tens of Tesla.

Here, we show that the ultrafast Barnett effect associated with circularly-polarized phonons can be exploited to achieve full-scale remote switching of magnetization. Our concept is shown schematically in Fig. 1a. Using circularly-polarized ultrashort light fields with mid-IR frequencies, we selectively drive at resonance phonons found not in the magnetic material of interest but rather in the paramagnetic substrate. The microscopic corkscrewing motion of the ions, via the Barnett effect, macroscopically induces spontaneous magnetization $\mathbf{M}_{BE}$ that is oriented normal to the plane of rotation. The direction of $\mathbf{M}_{BE}$ governs the direction of magnetic switching found in the nearby magnetic nanolayer, facilitating selective control over the final magnetic state. The fact that the switching is mediated by the substrate and is broadly unreliant on properties of the magnetic material of interest implies that the demonstrated method is very general, and could be potentially used to manipulate magnetic ordering on a universal scale.

In our experiments, we primarily study bilayers of 20-nm-thick GdFeCo and 5-nm-thick Si$_3$N$_4$ mounted on c-cut sapphire, glass-ceramic and silicon substrates (see Fig. 1b). The uniaxial magnetocrystalline anisotropy of GdFeCo allows magnetization to point along one of the two directions normal to the sample surface, which we visualize using magneto-optical microscopy. High-frequency phonons in the substrate are selectively excited at resonance using circularly-polarized narrowband IR pulses delivered by the free-electron laser facility FELIX[16]. The latter supplies "micropulses" - at a repetition rate of 25 MHz – within 8-µs-long bursts ("macropulses") at a rate of 10 Hz. The micropulses, with central wavelength $\lambda$ varying from 7 to 22 µm (wavenumber 455-1429 cm$^{-1}$), bandwidth tunable between 0.3-2.5%, and electric fields on the order of hundreds of kilovolts per centimeter, are focused to a spot of diameter ~150 µm on the GdFeCo film. Previous measurements have shown that the micropulses are Fourier-transform-limited[17], leading to durations between hundreds of femtoseconds and several picoseconds depending on the combined wavelength and bandwidth. We expose the sample to a variety of different excitations ranging from single micro- or macro-pulses to sweeping a train of macropulses across the sample surface at a fixed speed.

We initially expose the sapphire-mounted GdFeCo/Si$_3$N$_4$ bilayer to micropulses with the wavelength $\lambda = 21$ µm and duration $\tau \sim 3.3$ ps. Single micropulses of this duration, whether linearly- or circularly-polarized, always induce demagnetization, i.e., a randomized distribution of fine magnetic domains. If, however, we use a circularly-polarized macropulse containing ~200 pulses, we clearly observe that the demagnetized area becomes encircled by a distinct ring of switched magnetization (see Fig. 1c), depending on the combination of optical helicity and initial magnetic polarity. This ring is unaffected by subsequent macropulses with the same polarization. This helicity-dependent effect is made even clearer by sweeping the circularly-polarized macropulses across the sample (see Fig. 1d). Upon sweeping the laser from left to right, the left edge of the ring appears to unidirectionally switch all magnetic domains in its path, leaving a uniform trail of light or dark domain[18]. This trail is then rendered visible or invisible depending on the orientation of the starting magnetization.

Aiming to elucidate the role of the excitation frequency, we now tune the central wavelength of the circularly-polarized macropulses and sweep the pulses across the sample surface. By recording magneto-optical images both before and after the process of sweeping, we are able to calculate the net helicity-dependent switching efficiency $\varepsilon$ (see Methods and Extended Data Fig. S1). The spectral dependence of $\varepsilon$, plotted in Fig. 2a, reveals that the switching is clearly pronounced for pumping wavelengths between 14-17 µm and 21-22 µm. In this IR spectral range, it has long been known that sapphire features a rich spectrum of transverse optical (TO) and longitudinal optical (LO) phonon modes[19-21]. Using the imaginary part of the refractive index $\kappa$ [21], we overlay in Fig. 2a the optical absorption spectrum $d_{Al_2O_3}(\omega) = 4\pi\kappa_{Al_2O_3}/\lambda$ characteristic of sapphire. The latter, corresponding to the population of TO phonons found in the substrate, clearly correlates well with the switching efficiency of the adjacent GdFeCo film. Similar experiments performed for the same magnetic bilayer of GdFeCo/Si$_3$N$_4$ mounted on a glass-ceramic substrate reveal a similar correlation in switching efficiency with the substrate's absorption $d_{SiO_2}(\omega)$ [23], as shown in Fig. 2b. In striking contrast, however, the same experiments performed with the magnetic bilayer grown on a silicon substrate demonstrate a complete absence of helicity-dependent switching (see Extended Data Fig. S3). We note that silicon does not have any significant population of optical phonons in this spectral range.

Keeping the optical wavelength fixed at $\lambda = 21$ µm, we next adjust a variety of parameters to explore how the efficiency of helicity-dependent switching in the sapphire-mounted bilayer is affected. First, we find that the switching efficiency increases towards 100% when the sweeping speed is reduced, i.e., when the sample is exposed to more circularly-polarized micropulses (see Supplementary Information). In light of this, when considering the order-of-magnitude difference in sweeping speed, the sapphire-mounted sample offers dramatically better switching efficiency compared to its glass-ceramic-mounted counterpart. Second, the switching efficiency is rather independent of the incident fluence when above a certain threshold (see Fig. 3a). This stems from the Gaussian spatial distribution of the optically-delivered thermal load[19]. At the center of the irradiated area, the excess thermal energy completely destroys the magnetization for too long a time, thus randomizing its direction of recovery and resulting in a multi-domain structure. At the perimeter, in contrast, the thermal load is sufficiently low enough to lend control to the circular polarization, generally giving rise to a ring of switched magnetization. As shown by the magneto-optical images in the inset of Fig. 3a, the optical fluence only scales the size of this characteristic magnetic domain pattern, and thus our sweeping experiments broadly measure the switching isolated at the spot's perimeter. Third, as shown in Fig. 3b, varying the pulse duration $\tau$ from hundreds of femtoseconds to almost 4 ps leaves the switching efficiency broadly unaffected. This indifference to $\tau$ persists even when elevating the starting temperature of our experiments by 80 K (inset of Fig. 3b). Fourth, as shown in Fig. 3c, removing the $Si_3N_4$ interlayer dramatically diminishes the switching, but increasing its thickness from 5 nm to 50 nm preserves ~100% switching efficiency. While the spectral population of phonons in $Si_3N_4$ does not correlate at all with the helicity- and wavelength-dependent switching efficiency found in the magnetic nanolayer[24] (see Supplementary Information), the dielectric $Si_3N_4$ interlayer is clearly important for the switching process. Fifth and finally, the switching can be equally achieved by pumping macropulses with highly elliptical polarization. As shown in Fig. 3d, an ellipticity of just a few degrees is sufficient to successfully drive magnetic switching even with the $Si_3N_4$ interlayer being 50-nm thick. The mechanism is clearly robust.

Let us compare the discovered spectrally- and helicity-dependent magnetic switching shown in Figs. 1-2 to that achieved previously in ferromagnets using circularly-polarized light pulses with frequencies in the visible spectral range[30,31]. The root of the latter is still a controversial subject of debate, with the two potential mechanisms involving either the opto-magnetic inverse Faraday effect or magnetic circular dichroism associated with domain-wall expansion[32,33]. Neither of these drivers can account for the switching identified here, however. Magnetic circular dichroism depends on the helicity-dependent absorption spectrum of the magnetic layer, and should not hinge on the choice of substrate. Moreover, the absence of switching in the silicon-mounted bilayer proves that the IR-active phonons in the underlying substrate must be involved in the switching. Similarly, state-of-the-art *ab initio* models of the inverse Faraday effect in metals explicitly predict that the light-induced magnetization grows monotonically in strength with increasing wavelength[32]. This is in direct contradiction with our results shown in Fig. 2.

To explain our results, we argue that the circularly-polarized high-frequency phonons in the substrate, driven at resonance, play a vital role. Bulk sapphire ($\alpha$-$Al_2O_3$) crystallizes in the trigonal space group $R\bar{3}c$ with the conventional unit cell consisting of alternating hexagonal layers of $Al^{3+}$ cations and $O^{2-}$ anions stacked along the c-axis[19]. In our case, the latter is directed normal to the substrate plane. Sapphire features six IR-active phonon modes that are differentiated by their dipole-moments oscillating either parallel (two $A_{2u}$ modes) or perpendicular (four $E_u$ modes) to the c-axis[19]. There are further split in to TO and LO phonons. In our experiments, we illuminate the sample with IR pulses coming at normal incidence, and so the light can only couple to the four $E_u$ TO modes centered about 635 cm$^{-1}$, 570 cm$^{-1}$, 440 cm$^{-1}$ and 385 cm$^{-1}$. While the precise nature of the interconnected vibrational motion of the $Al^{3+}$ and $O^{2-}$ ions within these phonon modes is complex[34-36], the vibrations of the $O^{2-}$ anions constituting these TO phonon modes are free to oscillate along orthogonal axes within the plane of the substrate's surface[37,38] (as sketched in Fig. 1a). These doubly degenerate normal modes thus support the net excitation of a rotary vibration.

We can make a similar microscopic analysis of silica ($SiO_2$), which consists of tetrahedral units where four oxygen atoms surround the central silicon atom. Crystalline forms of silica feature optical phonons that are most commonly interpreted in terms of vibrations of oxygen atoms within the Si-O-Si subunits about the axis $\hat{z}$ passing through the two bridged Si atoms[39,40]. At the wavenumber about 450 cm$^{-1}$, the oxygen atom rocks about $\hat{z}$, whereas at 800 cm$^{-1}$, the oxygen atom symmetrically stretches perpendicular to $\hat{z}$. At 1100 cm$^{-1}$, the oxygen atom oscillates along $\hat{z}$ simultaneously stretching and compressing its adjacent Si-O bonds. Moreover, at this wavenumber, the oscillations of the O atoms within neighboring Si-O-Si subunits are in phase with each other.

We are now in a position to explain the helicity- and wavelength-dependent switching observed in Figs. 1-2. It is obvious that the circularly-polarized IR pulse thermally quenches the majority of the magnetization in the GdFeCo layer[1], but transfer-matrix-calculations show that the majority of the IR radiation actually transmits through the GdFeCo/$Si_3N_4$ bilayer to the surface of the substrate (see Supplementary Information). If appropriately tuned in frequency, the corkscrewing electric field of the optical pulse excites circularly-polarized TO phonons vibrating within the plane at the surface of the substrate. We emphasize that this does not depend on any symmetry-breaking intrinsic to the substrate since time-reversal symmetry is broken by the optical circular polarization. The oscillation of the optical phonons' dipole moment merely "follow" the rotating electric field of the IR light. Microscopically, the ensuing time-varying polarization $P$ of the optical phonon can polarize the paramagnetic spins directly via the cross-product $P \times \partial P/\partial t$ [41] or more indirectly via changes of the crystal field [14]. In either scenario, the orbital angular momentum of the rotational lattice vibration creates - for the lifetime of the phonon - a Barnett-effect-induced magnetization $M_{BE}$. Depending on the helicity of the driven phonons, $M_{BE}$ is oriented either parallel or antiparallel to the initial state of the magnetization in the nearby magnetic nanolayer. This polarity of $M_{BE}$ therefore controls the helicity dependence of the switching.

The substrate's phonon-induced magnetization $M_{BE}$ clearly interacts non-locally with the magnetic nanolayer, but how is this interaction mediated? Since the switching can be achieved across a 50-nm-thick $Si_3N_4$ interlayer, we can exclude any interfacial exchange interaction. Moreover, the fact that increasing the thickness of the $Si_3N_4$ interlayer from 5 nm to 50 nm leaves the switching entirely unaffected (see Fig. 3c) suggests the interaction is rather long-ranged, strongly resembling the expected action of a magnetic field. We estimate the latter to be on the order of a milli-Tesla, based on considerations of the domain-wall motion in GdFeCo (see Supplementary Information). A more quantitative estimate is significantly complicated by the fact that $M_{BE}$ is accompanied by the ultrafast thermally-induced demagnetization of the magnetic nanolayer. In the measurements shown here, such demagnetization is actually necessary for achieving the switching, since it reduces the coercive field of the magnetic layer, helping $\mathbf{M}_{BE}$ to grow the magnetization's recovery in a certain direction. However, as well as driving circularly-polarized optical phonons at resonance, the impinging IR light also generates a significant amount of heat at the substrate's surface. This thermal energy has the potential to demagnetize the magnetic layer for too long, counteracting the impact of $\mathbf{M}_{BE}$. The dielectric $Si_3N_4$ interlayer, therefore, plays an important role of thermally isolating the GdFeCo layer from the substrate's surface, protecting the magnetic nanolayer from such persistent demagnetization. Nevertheless, careful time-resolved measurements of helicity-dependent reorientation of magnetization in a non-absorbing dielectric magnet, grown at significant distance from e.g. a c-cut $α$-$Al_2O_3$ substrate, should, in principle, allow $|\mathbf{M}_{BE}|$ to be extracted[43].

We anticipate that the ultrafast phononic Barnett effect identified here can be found in a wide variety of substrates since rotational lattice vibrations are found universally in condensed phases of matter. Thus, it provides a handle for remotely steering magnetization in a wide range of systems. Our results in Fig. 2a not only highlight the important role of TO phonons in realizing the ultrafast Barnett effect but also reveal an intriguing feature associated with the frequency of the LO phonon modes. In particular, when using an excitation frequency of 520 cm$^{-1}$ to pump the sapphire-mounted sample, we consistently found that the switching efficiency became negative (Fig. 2a). This physically corresponds to the sense of helicity-dependent switching becoming inverted, i.e., left-handed circular polarization $σ^+$ switches downward-pointing $\mathbf{M}^↓$, rather than upward-pointing $\mathbf{M}^↑$ as observed at all other pump wavelengths [see Extended Data Fig. 1n compared to panels (j) or (p)]. This behavior could potentially be explained by the fact that, at frequencies close to 520 cm$^{-1}$, both the real and imaginary parts of sapphire's permittivity approach close to zero (see Supplementary Information). This so-called epsilon-near-zero regime[44] - outside of the reststrahlenband - can give rise to counter-intuitive effects such as negative refraction[45], often studied within the context of metamaterials[46] and metasurfaces[47]; in particular, four-wave mixing can result in a time-reversed wave[48], which could flip the helicity of the surface-localized optical wave and thus invert the sense of switching.

**References and Notes:**

1. Barnett, S. J. Magnetization by Rotation. *Phys. Rev.* **6**, 239-270 (1915).
2. Barnett, S. J. Magnetization and Rotation. *Am. J. Phys.* **16**, 140-147 (1948).
3. Beaurepaire, E., Merle, J.-C., Daunois, A. & Bigot, J. Y. Ultrafast Spin Dynamics in Ferromagnetic Nickel. *Phys. Rev. Lett.* **76**, 4250-4253 (1996).
4. Dornes, C. et al. The ultrafast Einstein–de Haas effect. *Nature* **565**, 209-212 (2019).
5. Tauchert, S. R. et al. Polarized phonons carry angular momentum in ultrafast demagnetization. *Nature* **602**, 73–77 (2022).
6. Mankowsky, R., Först, M. & Cavalleri, A. Non-equilibrium control of complex solids by nonlinear phononics. *Rep. Prog. Phys.* **79**, 064503 (2016).

7. Rini, M. et al. Control of the electronic phase of a manganite by mode-selective vibrational excitation. *Nature* **449**, 72-74 (2007).
8. Stupakiewicz, A. et al. Ultrafast phononic switching of magnetization. *Nat. Phys.* **17**, 489–492 (2021).
9. Zhu, H. et al. Observation of chiral phonons. *Science* **359**, 579-582 (2018).
10. Nova, T. F. et al. An effective magnetic field from optically driven phonons. *Nat. Phys.* **13**, 132–136 (2017).
11. Rebane, Yu T. Faraday effect produced in the residual ray region by the magnetic moment of an optical phonon in an ionic crystal. *Zh. Eksp. Teor. Fiz* **84**, 2323-2328 (1983).
12. Sasaki, R., Nii, Y. & Onose, Y. Magnetization control by angular momentum transfer from surface acoustic wave to ferromagnetic spin moments. *Nat. Commun.* **12**, 2599 (2021).
13. Juraschek, D. M., Narang, P. & Spaldin, N. A. Phono-magnetic analogs to opto-magnetic effects. *Phys. Rev. Res.* **2**, 043035 (2020).
14. Juraschek, D. M., Neuman, T. & Narang, P. Giant effective magnetic fields from optically driven chiral phonons in 4$f$ paramagnets. *Phys. Rev. Res.* **4**, 013129 (2022).
15. Juraschek, D. M. & Spaldin, N. A. Orbital magnetic moments of phonons. *Phys. Rev. Mater.* **3**, 064405 (2019).
16. Knippels, G. M. H. et al. Generation and Complete Electric-Field Characterization of Intense Ultrashort Tunable Far-Infrared Laser Pulses. *Phys. Rev. Lett.* **83**, 1578 (1999).
17. Knippels, G. M. H. & van der Meer, A. F. G. FEL diagnostics and user control, *Nucl. Instrum. Methods Phys. Res.* **144**, 32-39 (1998).
18. Khorsand, A. R. et al. Role of Magnetic Circular Dichroism in All-Optical Magnetic Recording. *Phys. Rev. Lett.* **108**, 127205 (2012).
19. Barker, Jr, A. S. Infrared Lattice Vibrations and Dielectric Dispersion in Corundum. *Phys. Rev.* **132**, 1474 (1963).
20. Gervais, F. and Piriou, B. Anharmonicity in several-polar-mode crystals: adjusting phonon self-energy of LO and TO modes in $Al_2O_3$ and $TiO_2$ to fit infrared reflectivity. *J. Phys. C: Solid State Phys.* **7**, 2374 (1974).
21. Schubert, M., Tiwald, T. E. and Herzinger, C. M. Infrared dielectric anisotropy and phonon modes of sapphire. *Phys. Rev. B* **61**, 8187 (2000).
22. Querry, M. R. *Optical constants. Contractor Report* CRDC-CR-85034 (1985).
23. Popova, S., Tolstykh, T. & Vorobev. V. Optical characteristics of amorphous quartz in the 1400–200 cm$^{-1}$ region, *Opt. Spectrosc.* **33**, 444–445 (1972).
24. Luke, K., Okawachi, Y., Lamont, M. R. E., Gaeta, A. L. & Lipson, M. Broadband mid-infrared frequency comb generation in a $Si_3N_4$ microresonator. *Opt. Lett.* **40**, 4823-4826 (2015).
25. Kimel, A. V., Kalashnikova, A. M., Pogrebna, A. & Zvezdin, A. K. Fundamentals and perspectives of ultrafast photoferroic recording. *Phys. Rep.* **852**, 1-46 (2020).
26. Ostler, T. A. et al. Ultrafast heating as a sufficient stimulus for magnetization reversal in a ferrimagnet. *Nat. Commun.* **3**, 666 (2012).
27. Mentink, J. H. et al. Ultrafast Spin Dynamics in Multisublattice Magnets. *Phys. Rev. Lett.* **108**, 057202 (2012).

28. Davies, C. S. et al. Pathways for Single-Shot All-Optical Switching of Magnetization in Ferrimagnets. *Phys. Rev. Appl.* **13**, 024064 (2020).
29. Davies, C. S. et al. Exchange-driven all-optical magnetic switching in compensated 3*d* ferrimagnets. *Phys. Rev. Res*. **2**, 032044(R) (2020).
30. Lambert, C.-H. et al. All-optical control of ferromagnetic thin films and nanostructures. *Science* **345**, 1337-1340 (2014).
31. Mangin, S. et al. Engineered materials for all-optical helicity-dependent magnetic switching. *Nat. Mater.* **13**, 286–292 (2014).
32. Berritta, M., Mondal, R., Carva, K. & Oppeneer, P. M. *Ab Initio* Theory of Coherent Laser-Induced Magnetization in Metals. *Phys. Rev. Lett.* **117**, 137203 (2016).
33. Quessab, Y. et al. Resolving the role of magnetic circular dichroism in multishot helicity-dependent all-optical switching. *Phys. Rev. B* **100**, 024425 (2019).
34. Bhagavantam, S. & Venkatarayudu, T. Raman effect in relation to crystal structure. *Proc. Natl. Acad. Sci. India A* **9**, 224-258 (1939).
35. Cowley, E. R. Symmetry properties of the normal modes of vibration of calcite and α-corundum. *Can. J. Phys.* **47**, 1381-1391 (1969).
36. Onari, S., Arai, T. & Kudo, K. Infrared lattice vibrations and dielectric dispersion in $\alpha-Fe_2O_3$. *Phys. Rev. B* **16**, 1717 (1977).
37. Sharma, A. & Singisetti, U. Low field electron transport in $\alpha$-$Ga_2O_3$: An ab initio approach. *Appl. Phys. Lett.* **118**, 032101 (2021).
38. Stokey, M. et al. Infrared dielectric functions and Brillouin zone center phonons of α-$Ga_2O_3$ compared to α-$Al_2O_3$. *Phys. Rev. Mater.* **6**, 014601 (2022).
39. Kirk, C. T. Quantitative analysis of the effect of disorder-induced mode coupling on infrared absorption in silica. *Phys. Rev. B* **38**, 1255 (1988).
40. Gunde, M. K. Vibrational modes in amorphous silicon dioxide. *Phys. B. Condens. Matter* **292**, 286-295 (2000).
41. Juraschek, D. M., Fechner, M., Balatsky, A. V. & Spaldin, N. A. Dynamical multiferroicity. *Phys. Rev. Mater.* **1**, 014401 (2017).
42. Zhang, L. & Niu, Q. Angular Momentum of Phonons and the Einstein–de Haas Effect. *Phys. Rev. Lett.* **112**, 085503 (2014).
43. Kimel, A. V. et al. Ultrafast nonthermal control of magnetization by instantaneous photomagnetic pulses. *Nature* **435**, 655-657 (2005).
44. Reshef, O., De Leon, I., Alam, M. Z. & Boyd, R. W. Nonlinear optical effects in epsilon-near-zero media. *Nat. Rev. Mater.* **4**, 535–551 (2019).
45. Kuroda, N., Tsugawa, K. & Yokoi, H. Negative Refraction of Infrared Waves and Rays in Sapphire α-$Al_2O_3$. *J. Phys. Soc. Jpn.* **81**, 114706 (2012).
46. Smith, D. R., Pendry, J. B. & Wiltshire, M. C. K. Metamaterials and Negative Refractive Index. *Science* **305**, 788-792 (2004).
47. Glybovski, S. B., Tretyakov, S. A., Belova, P. A., Kivshar, Yu. S. & Simovski, C. R. Metasurfaces: From microwaves to visible. *Phys. Rep.* **634**, 1-72 (2016).

48. Vezzoli, S. et al., Optical Time Reversal from Time-Dependent Epsilon-Near-Zero Media. *Phys. Rev. Lett.* **120**, 043902 (2018).

## ACKNOWLEDGEMENTS

We thank all technical staff at FELIX for technical support. We gratefully acknowledge the Nederlandse Organisatie voor Wetenschappelijk Onderzoek (NWO-I) for their financial contribution, including the support of the FELIX Laboratory. I. R. is grateful to M. Jeannin and N. Passler for their assistance with the 4×4 transfer matrix calculations code.

## AUTHOR CONTRIBUTIONS

C.S.D, A.V.K. and A.K. conceived the project. A.T. fabricated the samples. C.S.D. performed the magneto-optical imaging together with F.G.N.F, and C.S.D. processed the experimental results. I.R. performed transfer-matrix calculations. C.S.D and A.K. jointly discussed the result and wrote the manuscript with contributions from all authors.

## COMPETING INTERESTS

The authors declare no competing interests.

## DATA AND MATERIALS AVAILABILITY

All data shown in the main text and in the supplementary materials are available from the corresponding author upon reasonable request.

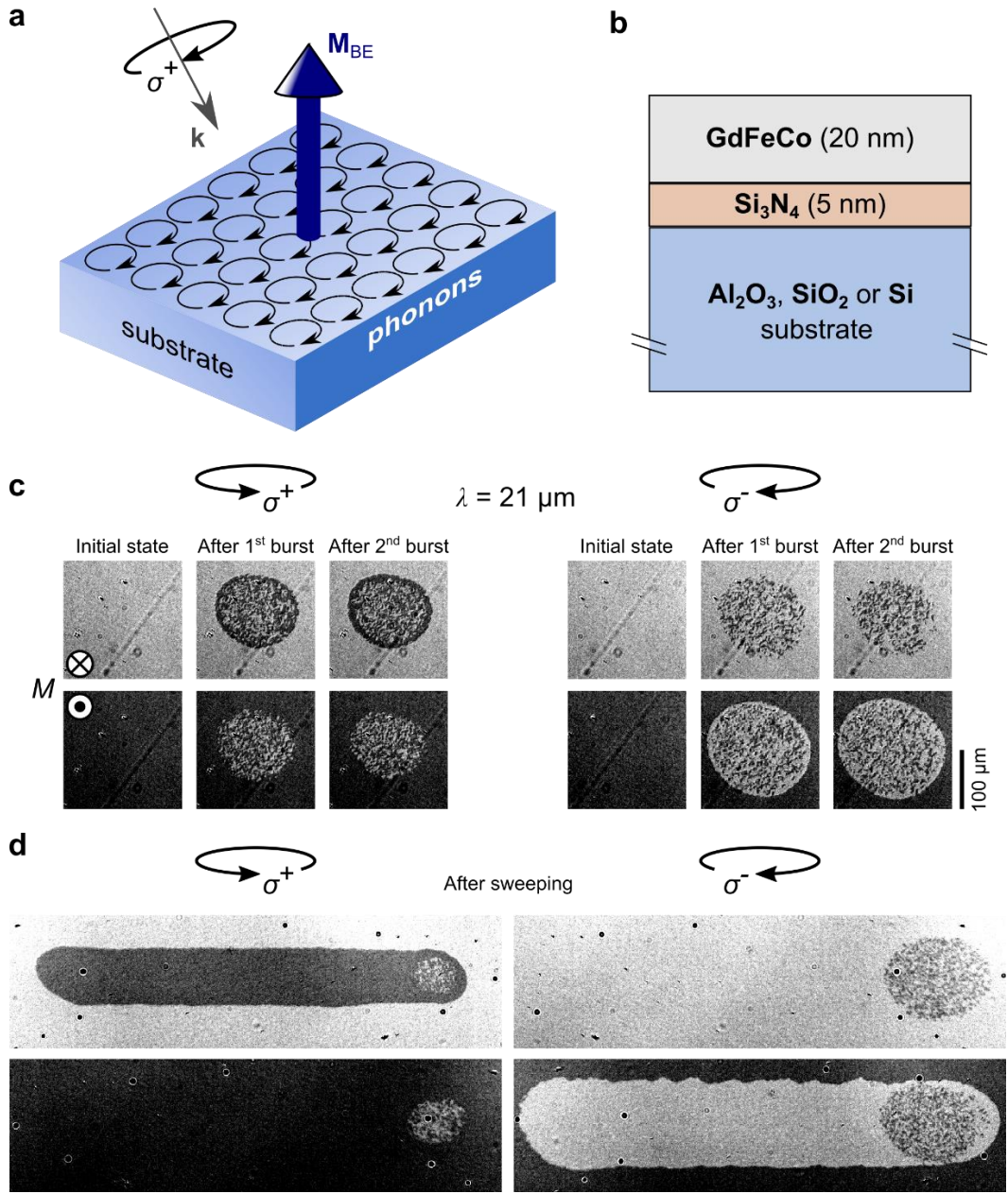

**Figure 1 | Concept of the phononic ultrafast Barnett effect and how it remotely switches magnetization. a** Sketch of how circularly-polarized optical phonons excited in a paramagnetic substrate, driven at resonance by circularly-polarized light fields with helicity $\sigma^+$, induces surface-localized magnetization $\mathbf{M}_{BE}$ via the Barnett effect. **b** The base heterostructure used in our experiments. **c** Magneto-optical images of the magnetization ($M$) of GdFeCo, mounted on a sapphire substrate, taken before (left panel) and after exposure to one (middle panel) and two (right panel) 8-µs-long macropulses. Each macropulse comprises a burst of circularly-polarized pulses with central wavelength $\lambda = 21$ µm and duration $\tau \sim 3.3$ ps coming at a rate of 25 MHz. **d** Result of continuously sweeping the macropulses from left to right at a speed of 20 µm/s across a single-domain background. The scale bar is common to all images shown in (c)-(d).

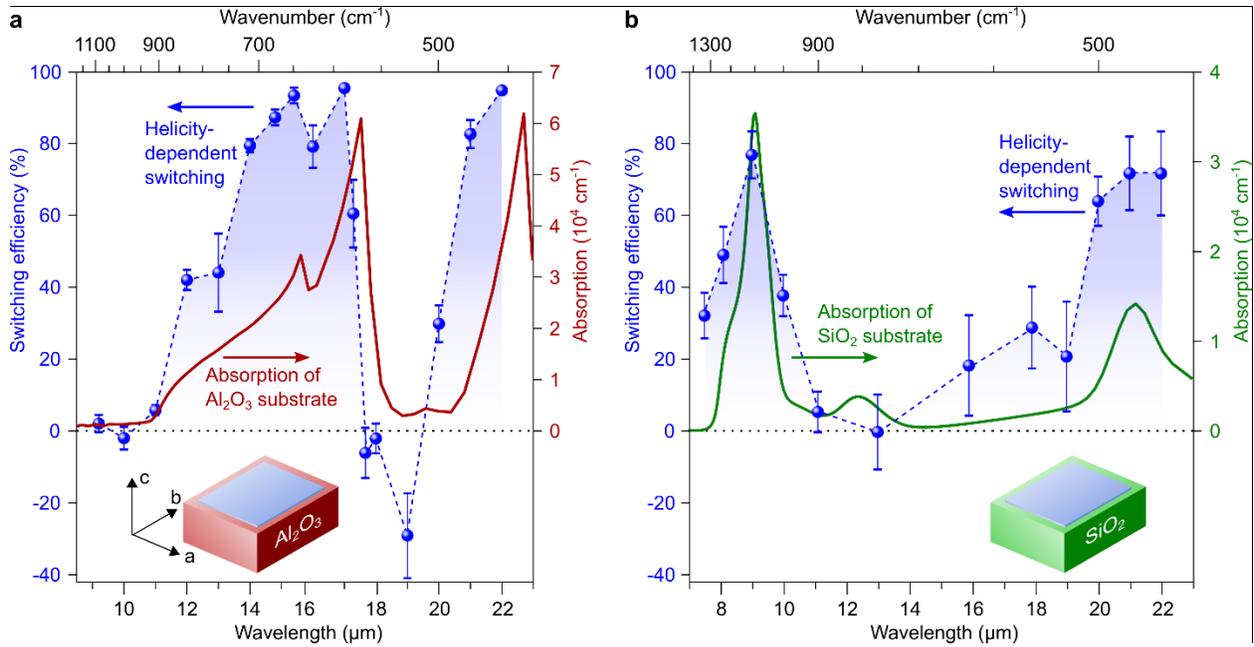

**Figure 2 | Resonant helicity-dependent switching of magnetization. a,b** Spectral dependence of the helicity-dependent switching of magnetization measured in the same GdFeCo/Si$_3$N$_4$ bilayer grown on a (a) sapphire and (b) glass-ceramic substrate, obtained with a sweeping speed of 50 µm/s and 5 µm/s respectively. Overlaid are the respective absorption spectra characteristic of the substrates. The net switching efficiency is defined as 0% when the GdFeCo film shows complete demagnetization irrespective of the optical helicity. The example images shown in Fig. 1d display a net switching efficiency of 100%.

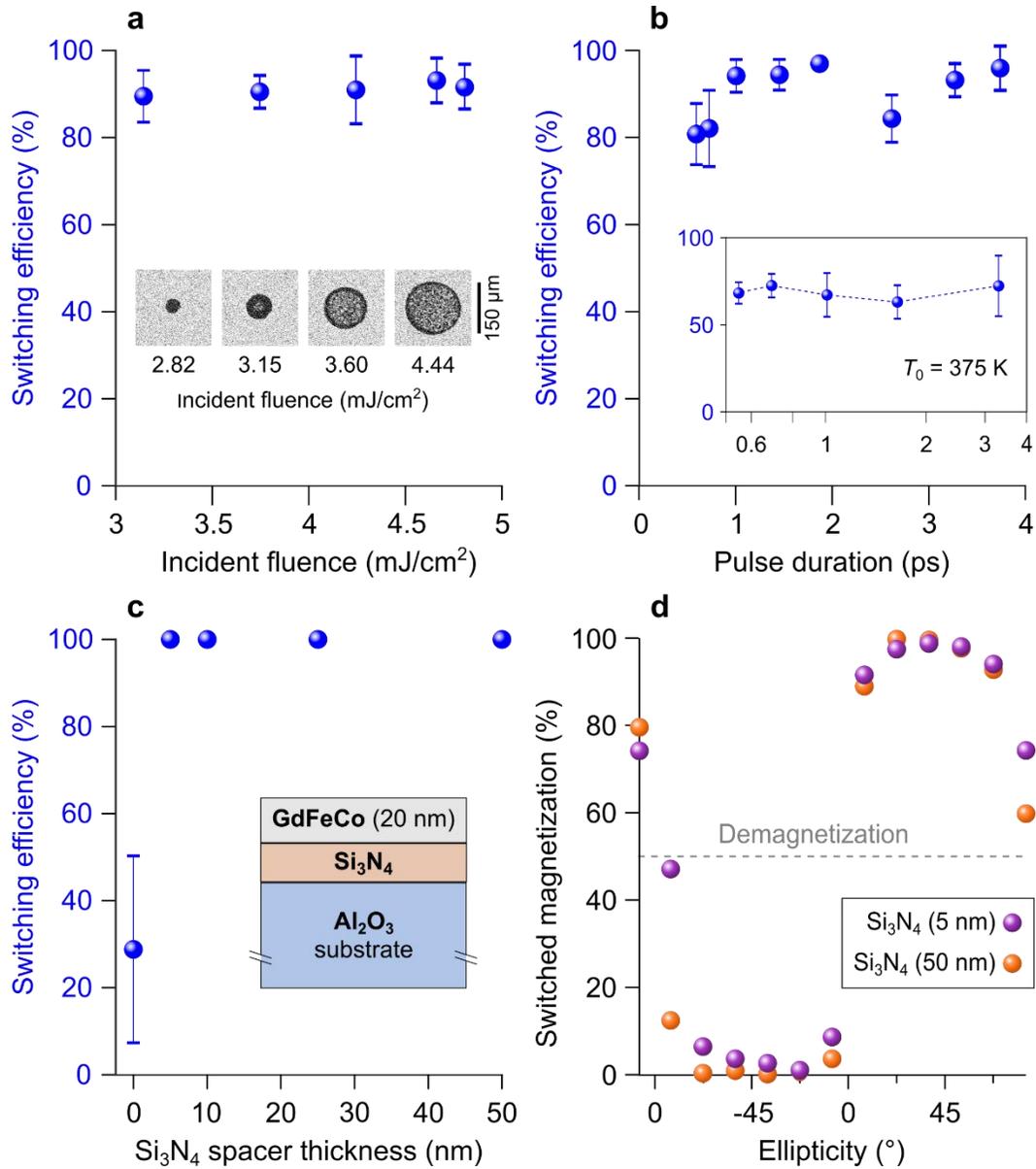

**Figure 3 | Dependence of switching effects on optical and interlayer parameters. a** Switching efficiency measured as a function of the incident fluence with a fixed pulse duration of ~3.3 ps and a sweeping speed of 20 µm/s. Inset: magneto-optical images taken after exposing the sapphire-mounted sample to a single circularly-polarized macropulse. **b** Switching efficiency measured as a function of the pulse duration, obtained at room temperature with a sweeping speed of 10 µm/s. Inset: same as in the main panel but with the sample's starting temperature $T_0$ raised to 375 K. **c** Switching efficiency measured for the same sapphire-mounted heterostructure (shown schematically in the inset) with varying thickness of the interfacial $Si_3N_4$ layer, measured with a sweeping speed of 5 µm/s. **d** Magnetization switched by macropulses with varying ellipticity, using samples with a 5-nm- and 50-nm-thick interfacial $Si_3N_4$ layer and a sweeping speed of 20 µm/s. In panels (a)-(c), the central wavelength of the pump pulse was $\lambda = 21$ µm, whereas in panel (d), $\lambda = 16.2$ µm.

**METHODS**

**Materials**

The studied samples were 20-nm-thick $Gd_{24}Fe_{66.5}Co_{9.5}$ amorphous films, grown on a $Si_3N_4$ buffer layer of thickness $x$ ($x$ = 0, 5, 10, 25 or 50 nm), deposited by DC magnetron sputtering. To protect the GdFeCo layer and prevent oxidation, a 60-nm-thick capping layer of $Si_3N_4$ layer is always used. The nanostructure was grown on one of three substrates. The first is a double-side-polished (0001) c-cut sapphire substrate, commercially obtained either from Alineason or Ossila. This was grown by the Kyropolous technique and has purity >99.9% and thickness 500 µm (Alineason) or 1.1 mm (Ossila). The second is an atomically flat glass-ceramic substrate (serial TS-10 SX [49]), commercially obtained from Ohara, of thickness 675 µm and with a nominal surface roughness Ra ~ 0.6 Å. The third is a 500-µm-thick prime grade silicon substrate (double-side polished) commercially obtained from Prime Wafers.

**Methods**

*A. Magneto-optical microscopy experiments at FELIX*

To detect how the magnetization of the GdFeCo films are affected by IR pulses, we use the technique of magneto-optical polarized microscopy. A LED lamp illuminates the sample at normal incidence with linearly-polarized white light. The white light transmitted through the sample is gathered by an objective lens, filtered by an analyzer and imaged using a CCD camera. The GdFeCo film has uniaxial magnetocrystalline anisotropy directed normal to the sample plane. Thus, via the magneto-optical Faraday effect, the axis of the light's linear polarization rotates in opposite directions depending on whether the magnetization points parallel or antiparallel to the sample normal. The filtering by the analyzer therefore visualizes the two different equilibrium magnetic domains as light or dark. With a typical exposure time on the order of 100 ms, we record the magnetic domains after illumination by the infrared pulses and in the two saturated states. Optionally, the sample is mounted on a resistive heater allowing the sample's ambient temperature $T_0$ to be raised from room temperature to 400 K.

The IR pulses delivered by FELIX (Free Electron Lasers for Infrared eXperiments) are tuned to a central wavelength in the spectral range 8-22 µm with a bandwidth experimentally adjustable in the range 0.35−2.5%. Since the pulses are Fourier-transform-limited, we estimate the wavelength-dependent pulse duration using the measured spectrum obtained *in-situ* by splitting a fraction of the IR beam. A KRS-5 wire grid polarizer assures linear polarization of the IR pulse, and a tunable zero-order CdSe quarter waveplate (Alphalas) provides circular polarization. The transmission of CdSe restricted the maximum usable wavelength to 22 µm. The IR pulses are focused using a 90° off-axis parabolic mirror (focal-length 75.6 mm) on to the sample surface, to a spot with diameter ≈150 µm. The parabolic mirror is mounted on a motorized translation axis, allowing the IR pulses to be swept across the sample surface at a fixed speed with no change in the spot profile. Further details of the IR pulses are provided in the Supplementary Information.

*B. Post-processing of gathered magneto-optical images*

During the magneto-optical microscopy experiments, we consistently record an image after sweeping IR pulses across the sample surface, and a reference image taken after saturating the magnetization of the GdFeCo layer parallel to its original polarity. These images are hereafter referred to as $I$ and $M^{sat}$ respectively. To extract the switching efficiency $\varepsilon$, we first inspect image $I$ and average the magneto-optical intensity across a rectangular region spanning the track of the swept IR pulse. The net switched magnetization $\langle M \rangle(\sigma^+, M^\uparrow)$, obtained with the optical helicity $\sigma^+$ and starting magnetic polarity $M^\uparrow$, is then calculated by normalizing the aforementioned average with that similarly obtained from the reference image $M^{sat}$. When $\langle M \rangle = 0\%$ (100%), the magnetization is completely unswitched (switched), whereas $\langle M \rangle = 50\%$ corresponds to complete demagnetization.

Second, this protocol is used to analyze all four images with different optical helicities and starting magnetic polarities, yielding four scalar values $\langle M \rangle(\sigma^+, M^\uparrow)$, $\langle M \rangle(\sigma^+, M^\downarrow)$, $\langle M \rangle(\sigma^-, M^\uparrow)$ and $\langle M \rangle(\sigma^-, M^\downarrow)$.

Third, we calculate the switching efficiency $\varepsilon$ as
$$\varepsilon = [\langle M \rangle(\sigma^+, M^\uparrow) - \langle M \rangle(\sigma^-, M^\uparrow) - \langle M \rangle(\sigma^+, M^\downarrow) + \langle M \rangle(\sigma^-, M^\downarrow)] / 4.$$
This produces $\varepsilon = 0$ when the laser pulses demagnetize the sample regardless of the optical helicity and initial magnetic polarity. On the contrary, $\varepsilon = 100\%$ when circularly-polarized pulses of helicity $\sigma^+$ and $\sigma^-$ uniformly switch magnetization $M^\uparrow$ to $M^\downarrow$ and $M^\downarrow$ to $M^\uparrow$ respectively, but not vice versa. This corresponds to the magneto-optical images shown in Fig. 1(d).

The switching efficiency $\varepsilon$ can becomes negative if the sense of switching flips, i.e., circularly-polarized pulses of helicity $\sigma^+$ and $\sigma^-$ now switch magnetization $M^\downarrow$ to $M^\uparrow$ and $M^\downarrow$ to $M^\uparrow$ respectively. This is clearly observed in our experimental measurements shown in Extended Data Fig. 1(n). In this panel, the top-right and bottom-left subpanels show better switching compared to the top-left and bottom-right ones. This is opposite in sense to the switching observed at other wavelengths, e.g., in Extended Data Fig. 1(o), the top-left / bottom-right subpanels show better switching than that observed in the top-right / bottom-left ones.

The error bars in Figs. 2-3 correspond to ±1 standard deviation of the four scalar quantities $\langle M \rangle(\sigma^+, M^\uparrow)$, $[100 - \langle M \rangle(\sigma^-, M^\uparrow)]$, $[100 - \langle M \rangle(\sigma^+, M^\downarrow)]$ and $\langle M \rangle(\sigma^-, M^\downarrow)$.

*C. Infrared reflectivity measurements*

To measure the reflectivity of the studied samples in the IR spectral range, we use the narrow-band pulses delivered by FELIX. A 5-mm-thick KRS-5 window inclined at an angle of ~20° splits the IR pulse in two, reflecting ~15% and transmitting ~70% independent of the pulse's wavelength. The reflected IR macropulse is detected by a $N_2$-cooled HgCdTe detector (InfraRed Associates), while the transmitted part is focused onto the sample at an angle close to normal incidence (~5°). The beam reflected from the sample is subsequently measured using a second $N_2$-cooled HgCdTe detector. The ratio between the two detected signals then yields the sample reflectivity, compensating for the intensity fluctuations associated with the IR laser. The measurements were performed at room temperature in atmospheric conditions, similar to the experimental measurements performed throughout. The results presented in Supplementary Note 2 were obtained by averaging over 4 individual measurements.

*D. Transfer-matrix calculations*

To calculate the distribution of the electric field across the multilayered nanostructure, we use the 4×4 transfer matrix method[50] recently generalized and optimized[51]. We consider a structure of composition $Si_3N_4(60)/Fe(20)/Si_3N_4(5)/glass(5\times10^5)$, where the number in parentheses corresponds to the layer thickness in nanometers, using the wavelength-dependent complex refractive indices $n' = n + i\kappa$ presented in Refs. [23,24]. In lieu of accurate knowledge of $n'_{GdFeCo}$ in the infrared spectral range, we use values typical of iron[22]. This assumption is justified by the reflectivity being broadly unaffected by substitution of the GdFeCo layer with iron (see

Supplementary Information). Using the procedure outlined in Ref. [51], we evaluate the elements of the Fresnel reflection tensor for each layer. The total absorption $A$ can be extracted as 1-$R$-$T$ where $R$ and $T$ is the reflection and transmission respectively. Alternatively, we estimate the layer-specific $A$ as integrated squares of the absolute values of the electric field $E$ inside each layer. Both of these methods produce spectral shapes of total $A$, integrated over all layers, that is proportional to each other. The step in the direction normal to the sample plane was small enough (0.1 nm) so that further reduction had no impact on the results of the calculations.

**Supplementary References and Notes:**

49. Goto, N. et al., U.S. Patent Number 5,391,522, February 21, 1995
50. Born, M. & Wolf, E. *Principles of Optics* (Pergamon, Oxford, 1980).
51. Passler, N. C. & Paarmann, A. Generalized 4 × 4 matrix formalism for light propagation in anisotropic stratified media: Study of surface phonon polaritons in polar dielectric heterostructures. *J. Opt. Soc. Am. B* **34**, 2128-2139 (2017).
52. Nilsson, G. & Nelin, G. Phonon Dispersion Relations in Ge at 80 °K. *Phys. Rev. B* **3**, 364 (1971).
53. Jhansirani, K., Dubey, R. S., More, M. A. & Singh, S. Deposition of silicon nitride films using chemical vapor deposition for photovoltaic applications. *Results Phys.* **6**, 1059-1063 (2016).
54. Parsons, G. N., Souk, J. H. & Batey, J. Low hydrogen content stoichiometric silicon nitride films deposited by plasma-enhanced chemical vapor deposition. *J. Appl. Phys.* **70**, 1553 (1991).
55. Davies, C. S., Mentink, J. H., Kimel, A. V., Rasing, Th. & Kirilyuk, A. Helicity-independent all-optical switching of magnetization in ferrimagnetic alloys. *J. Magn. Magn. Mater.* **563**, 169851 (2022).
56. Quessab, Y. et al., Resolving the role of magnetic circular dichroism in multishot helicity-dependent all-optical switching. *Phys. Rev. B* **100**, 024425 (2019).
57. Stupakiewicz, A., Szerenos, K., Afanasiev, D., Kirilyuk, A. & Kimel, A. V. Ultrafast nonthermal photo-magnetic recording in a transparent medium. *Nature* **542**, 71–74 (2017).
58. Stupakiewicz, A. et al., Ultrafast phononic switching of magnetization. *Nat. Phys.* **17**, 489–492 (2021).
59. Knippels, G. M. H. et al., Two-color facility based on a broadly tunable infrared free-electron laser and a subpicosecond-synchronized 10-fs-Ti:sapphire laser. *Opt. Lett.* **23**, 1754-1756 (1998).
60. Lisitsa, M. P. et al., Dispersion of the Refractive Indices and Birefringence of $CdS_xSe_{1-x}$ Single Crystals. *Phys. Stat. Sol.* **31**, 389 (1969).
61. Kim, K. J., Kim, S., Hirata, Y. et al. Fast domain wall motion in the vicinity of the angular momentum compensation temperature of ferrimagnets. Nature Mater **16**, 1187–1192 (2017).
62. Hellman, F., Abarra, E. N., Shapiro, A. L. & van Dover, R. B., Specific heat of amorphous rare-earth–transition-metal films. *Phys. Rev. B* **58**, 5672 (1998).
63. Pendry, J. Time Reversal and Negative Refraction. *Science* **322**, 71-73 (2008).
64. Kinsey, N. et al. Near-zero-index materials for photonics. *Nat. Rev. Mater.* **4**, 742–760 (2019).
65. Forati, E., Hanson, G. W. & Sievenpiper, D. F. An Epsilon-Near-Zero Total-Internal-Reflection Metamaterial Antenna. *IEEE Trans. Antennas Propag.* 6**3**, 1909-1916 (2015).

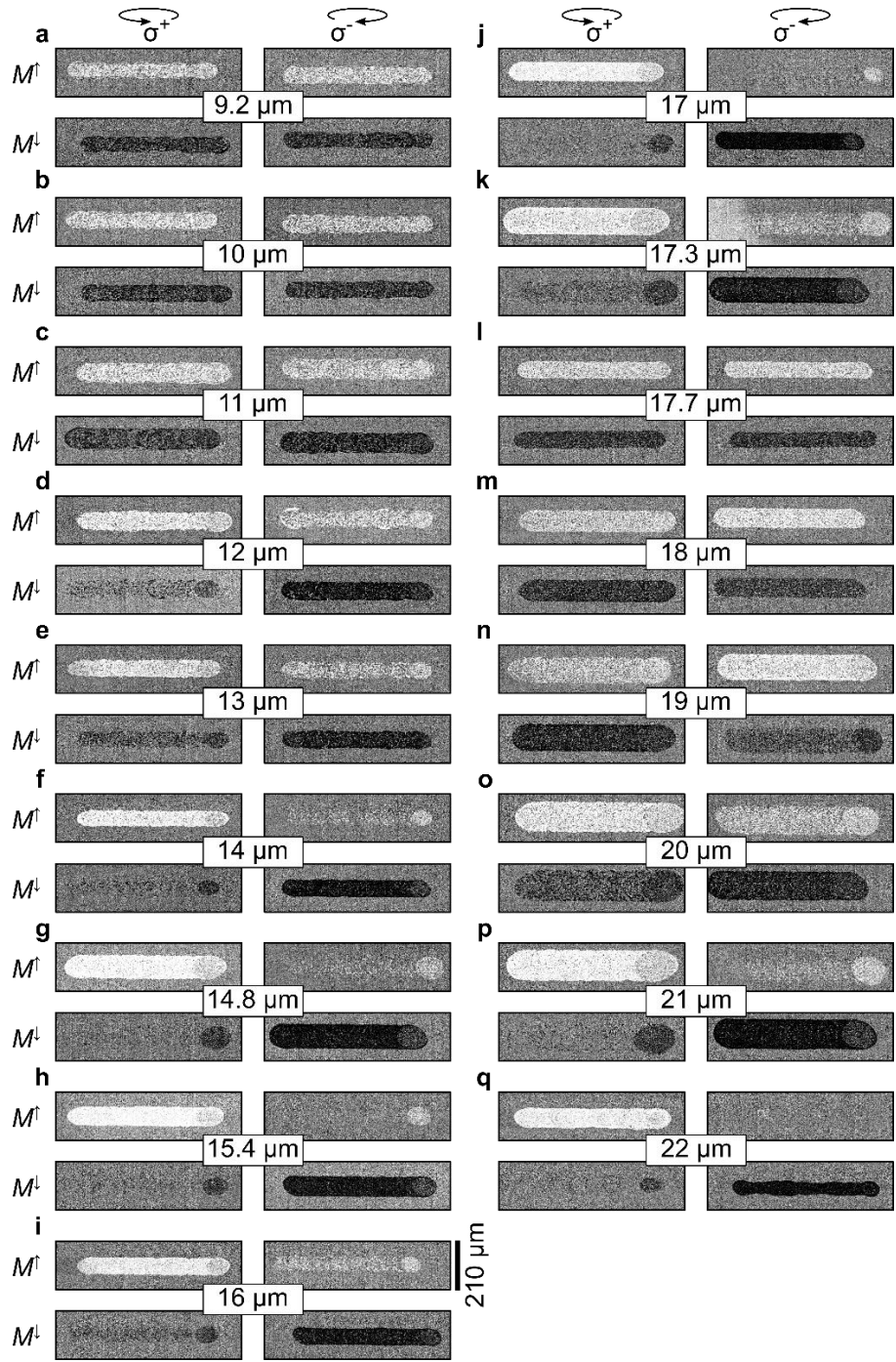

**Extended Data Fig. 1 | a-q** Background-subtracted magneto-optical images of the magnetization of GdFeCo, mounted on a sapphire substrate, taken after sweeping circularly-polarized "macropulses" with central wavelength $\lambda$ as indicated. The pulses, of helicity $\sigma^{\pm}$, are swept from left to right at a speed of 50 µm/s across a single-domain background with initial magnetic polarity $M^{\uparrow}$ or $M^{\downarrow}$. The scale bar is common to all images.

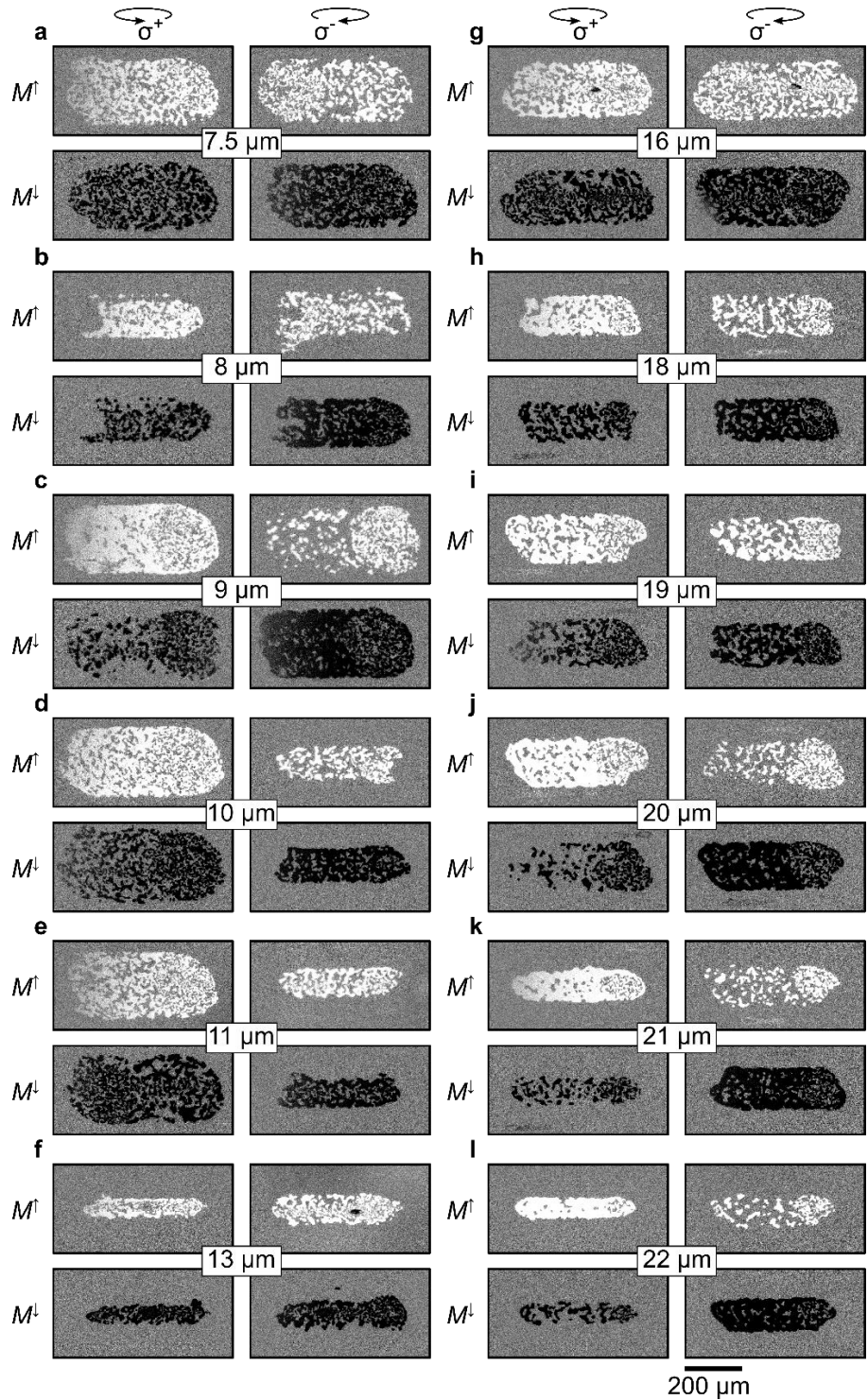

**Extended Data Fig. 2 | a-l** Background-subtracted magneto-optical images of the magnetization of GdFeCo, mounted on a glass-ceramic substrate, taken after sweeping circularly-polarized "macropulses" with central wavelength $\lambda$ as indicated. The pulses, of helicity $\sigma^{\pm}$, are swept from left to right at a speed of 5 µm/s across a single-domain background with initial magnetic polarity $M^{\uparrow}$ or $M^{\downarrow}$. The scale bar is common to all images.

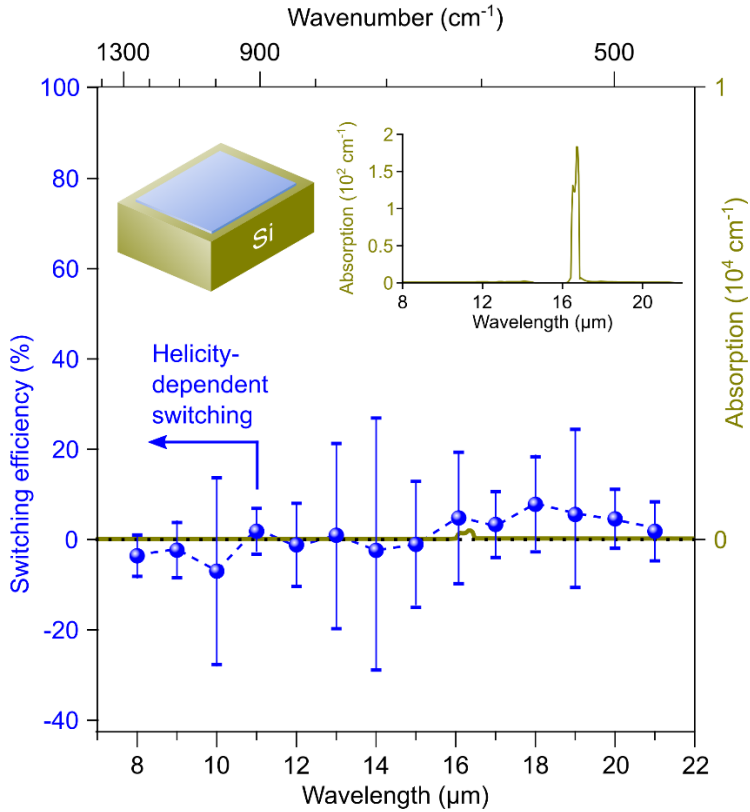

**Extended Data Fig. 3 |** Spectral dependence of the helicity-dependent switching of magnetization measured in a GdFeCo/Si$_3$N$_4$ bilayer grown on a silicon substrate, obtained with a sweeping speed of 40 µm/s. Overlaid is the respective absorption spectrum characteristic of the silicon substrate, with the inset providing a zoomed section. The net switching efficiency is defined as 0% when the GdFeCo film shows complete demagnetization irrespective of the optical helicity.

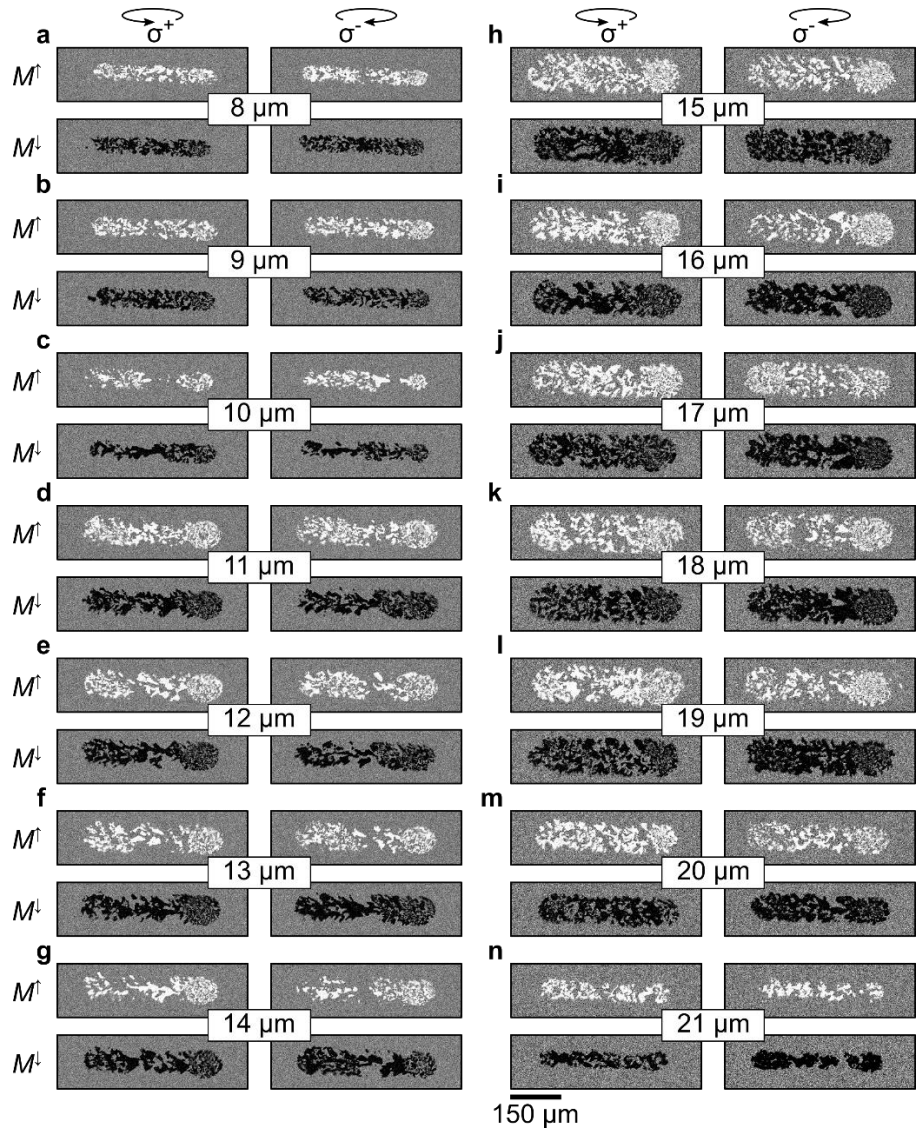

**Extended Data Fig. 4 / a-n** Background-subtracted magneto-optical images of the magnetization of GdFeCo, mounted on a silicon substrate, taken after sweeping circularly-polarized "macropulses" with central wavelength $\lambda$ as indicated. The pulses, of helicity $\sigma^{\pm}$, are swept from left to right at a speed of 40 µm/s across a single-domain background with initial magnetic polarity $M^{\uparrow}$ or $M^{\downarrow}$. The scale bar is common to all images.

SUPPLEMENTARY INFORMATION

# Phononic Switching of Magnetization by the Ultrafast Barnett Effect

C. S. Davies[1,2*], F. G. N. Fennema[1,2], A. Tsukamoto[3], I. Razdolski[1,2,4], A. V. Kimel[2] and A. Kirilyuk[1,2]

[1]FELIX Laboratory, Radboud University, Toernooiveld 7, 6525 ED Nijmegen, The Netherlands
[2]Radboud University, Institute for Molecules and Materials, Heyendaalseweg 135, 6525 AJ Nijmegen, The Netherlands
[3]College of Science and Technology, Nihon University, 7-24-1 Funabashi, Chiba 274-8501, Japan
[4]Faculty of Physics, University of Bialystok, Ciolkowskiego 1L, 15-245 Bialystok, Poland

*Correspondence to: carl.davies@ru.nl (C.S.D)

## Supplementary Note 1: Phonon modes and permittivity of c-cut sapphire

Bulk sapphire ($\alpha$-Al$_2$O$_3$) crystallizes in the trigonal space group $R\bar{3}c$ with the conventional unit cell consisting of alternating hexagonal layers of Al$^{3+}$ cations and O$^{2-}$ anions stacked along the c-axis. By consideration of the crystal's group symmetry, four distinct IR-active $E_u$ phonon modes should exist with a dipole moment oscillating perpendicular to the c-axis (electric field vector $E \perp c$), and two distinct $A_{2u}$ phonon modes with dipole moment oscillating parallel to the c-axis (electric field vector $E \parallel c$). The coulomb interaction splits all these phonon modes in to longitudinal and transverse optical ones (LO and TO respectively).

The optical phonon modes of sapphire have been extensively characterized by reflectivity and ellipsometry measurements performed over the course of more than half a century. In Supplementary Table 1, we present typical spectral positions of the TO and LO optical phonons belonging to $\alpha$-Al$_2$O$_3$ measured by three different research groups. The identified frequencies of phonon modes are all consistent with each other.

| Mode (#) | Barker (Ref. [S1]) TO | LO | Gervais et al. (Ref. [S2]) TO | LO | Schubert et al. (Ref. [S3]) TO | LO |
|---|---|---|---|---|---|---|
| $E_u$ (1) | 15.7 | 11.1 | 15.7 | 11.0 | 15.8 | 11.0 |
| $E_u$ (2) | 17.6 | 16.0 | 17.6 | 15.9 | 17.6 | 15.9 |
| $E_u$ (3) | 22.6 | 20.8 | 22.8 | 20.7 | 22.8 | 20.8 |
| $E_u$ (4) | 26.0 | 25.8 | 26.0 | 25.8 | 26.0 | 25.8 |
| $A_{2u}$ (1) | 17.2 | 11.5 | 17.1 | 11.3 | 17.2 | 11.3 |
| $A_{2u}$ (2) | 25.0 | 19.5 | 25.0 | 19.5 | 25.2 | 19.6 |

**Supplementary Table 1** | Central wavelengths (in units of µm, rounded to the nearest 0.1 µm) of the transverse and longitudinal optical phonon modes (TO and LO respectively) characteristic of $\alpha$-Al$_2$O$_3$, as obtained by Barker in Ref. [S1], Gervais et al. in Ref. [S2] and Schubert et al. in Ref. [S3].

Using the results shown in Supplementary Table 1, in combination with the associated linewidths and static/high-frequency dielectric constants, one can calculate the permittivity $\varepsilon = \varepsilon_1 + i\varepsilon_2$ by assuming the phonons are harmonic oscillators. Figure S1 shows the corresponding real and imaginary parts of sapphire's permittivity as extracted using the ellipsometry-based measurements presented by Schubert et al. in Ref. [S3]. The panels in column (a) and (b) correspond to the scenarios where $E \perp c$ and $E \parallel c$ respectively. The TO-phonon modes are indicated by maxima in $\varepsilon_2$.

In our measurements, we deliberately opted to study a c-cut sapphire substrate, with the c-axis perpendicular to the substrate's plane. Thus, since we illuminate the sample with IR pulses at normal incidence, we can directly couple to and drive the four $E_u$ modes. Taking in to account the paramagnetic nature of sapphire, we obtain the complex refractive index $n´ = n + i\kappa$ of sapphire via the relationship $n´ = \varepsilon^{½}$.

The phonon modes indicated in Supplementary Table 1 have an associated broadening $\Gamma_{FWHM}$, which defines the lifetime $\tau$ of the mode via the relationship $\tau^{-1} = \pi \cdot \Gamma_{FWHM}$ [S4]. Based on ellipsometry measurements of $\alpha$-$Al_2O_3$ performed at room temperature in equilibrium conditions, Schubert *et al.* in Ref. [S3] calculate that the TO $E_u$ phonon modes (1), (2), (3) and (4) referred to in Supplementary Table 1 have a Lorentzian broadening of 5.0 cm$^{-1}$, 4.7 cm$^{-1}$, 3.1 cm$^{-1}$ and 3.3 cm$^{-1}$ respectively. These nominally correspond to phonon lifetimes of 2.1 ps, 2.2 ps, 3.4 ps and 3.2 ps respectively. These lifetimes explain why we find that the switching is broadly independent of the pulse width (Fig. 3b), since our pulses have a duration that is shorter or equal to this lifetime.

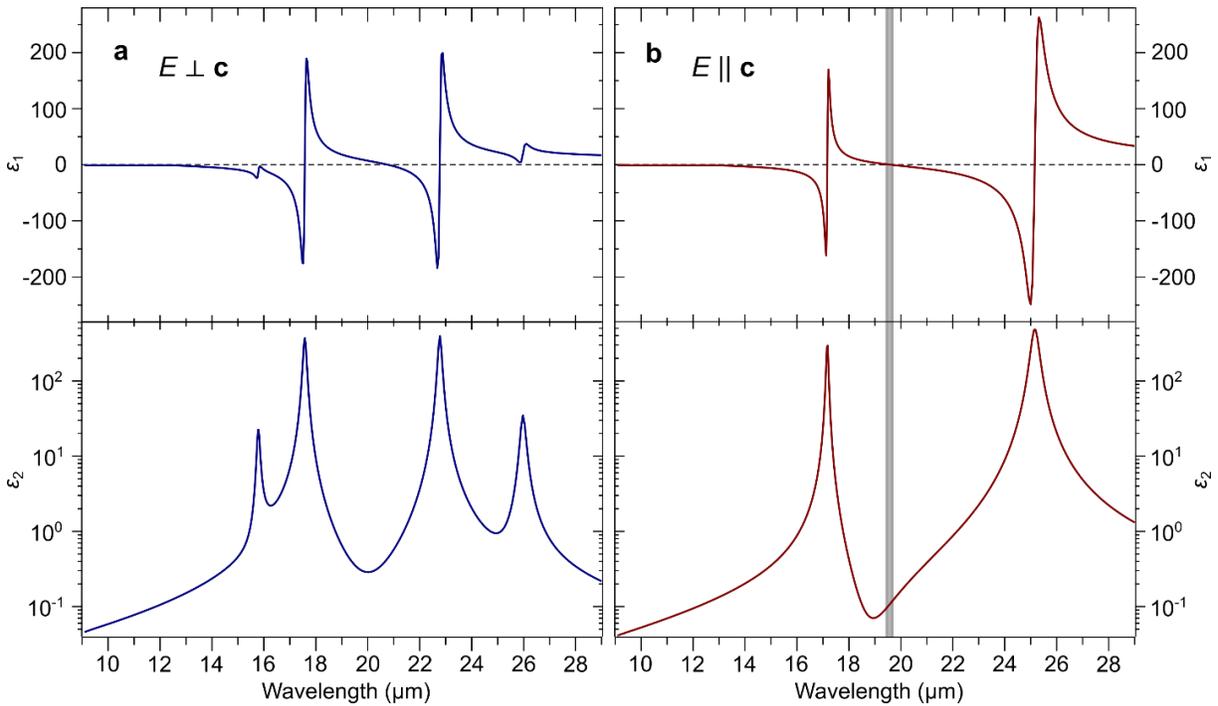

**Fig. S1 | a,b** Spectral dependence of the real and imaginary parts of the permittivity $\varepsilon = \varepsilon_1 + i\varepsilon_2$ of $\alpha$-$Al_2O_3$ (top and bottom panels respectively), calculated using the results of ellipsometry measurements presented by Schubert *et al.* in Ref. [S3]. The spectrum shown in columns (a) and (b) were measured with the electric field $E$ perpendicular and parallel to the c-axis of $\alpha$-$Al_2O_3$ respectively. The gray area shaded in panel (b) is discussed in Supplementary Note 10.

## Supplementary Note 2: Reflectivity of samples and substrates

In Fig. 2, we use measurements obtained from literature to characterize the optical absorption of the substrates in the IR spectral range. To assess the validity of this assumption, we present in Fig. S2 the measured reflectivity of the samples used in this study. In all panels, the points correspond to our measurements. The dashed blue lines in panels (a) and (b) correspond to the reflectivity deriving from the optical constants obtained by Schubert *et al.* (Ref. [S3]) and Popova *et al.* (Ref. [S5]) respectively. The smoothed solid lines in panels (c)-(d) provide a guide to the eye.

On one hand, the reflectivity of our c-cut sapphire substrate (Fig. 2a) matches well the spectrum presented in the literature. On the other hand, while our measurements shown in Fig. 2b broadly reproduce the strong reflectivity of quartz around the wavelengths 9 µm and 21 µm, our glass-ceramic substrate features additional small peaks about 16 µm and 18 µm. These features are attributed to the fact that our glass substrate, while predominantly containing 65-83% quartz ($\alpha$-$SiO_2$), also contains other constituents including 8-13% $Li_2Si_2O_5$, 0.5-5.5% MgO and 0-7% $K_2O$ (Ref. [S6]). Nevertheless, these features are substantially (>75%) weaker than the aforementioned peaks characteristic of quartz.

In Fig. S2c, we present the measured reflectivity of the sapphire-mounted GdFeCo/$Si_3N_4$ bilayer. We observe that the reflectivity is broadly flat between $8 \leq \lambda \leq 14$ µm, and then increases at wavelengths greater than 14 µm. Note that this neither correlates with the reflectivity spectrum characteristic of sapphire (Fig. S2a) nor with the spectral dependence of switching shown in Fig. 2a. In Fig. S2d, we show the reflectivity of glass-mounted GdFeCo/$Si_3N_4$ (black) and Fe/$Si_3N_4$ (gold). These structures are nominally identical save for the substitution of GdFeCo with Fe (both of thickness 20 nm). We note again that these reflectivity trends do not correlate with the switching shown in Fig. 2b in the main text.

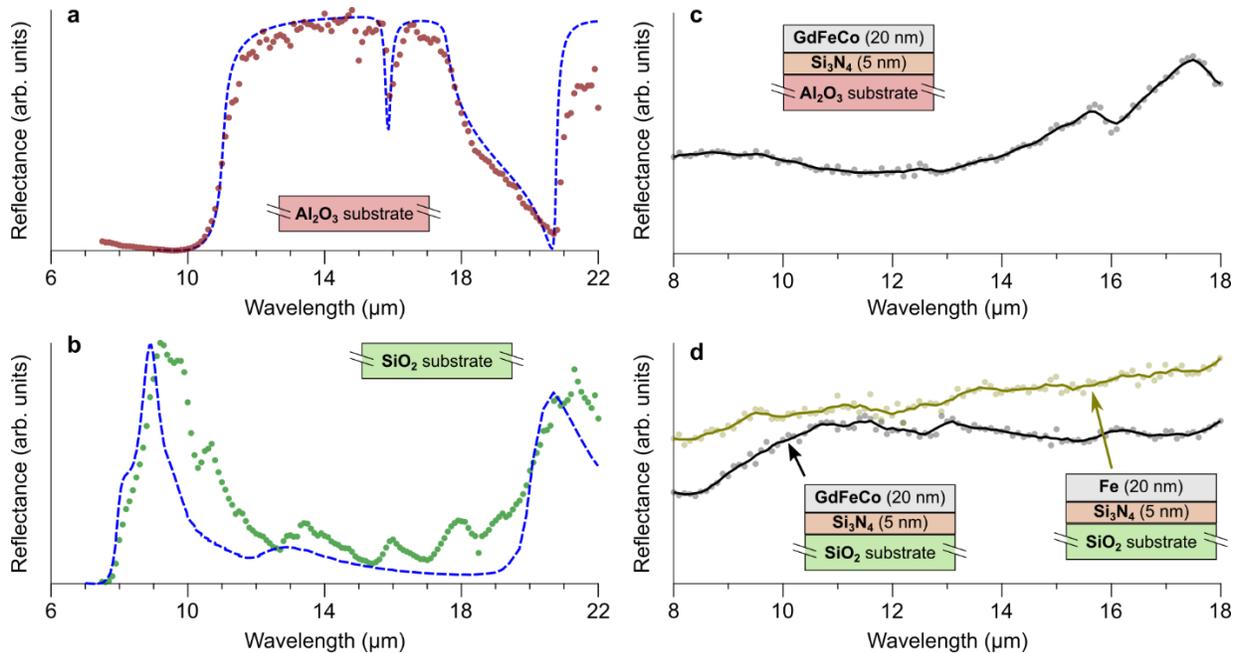

**Fig. S2 | a-b** The wavelength-dependent reflectivity of the sapphire and glass-ceramic substrate respectively (points). Overlaid is the expected reflectivity (blue dashed line) using the literature values presented in Ref. [S3] and Ref. [S5]. **c** Measured reflectivity of the sapphire-mounted GdFeCo/$Si_3N_4$ bilayer. **d** Measured reflectivity of the indicated glass-mounted heterostructures.

## Supplementary Note 3: Optical phonons and absorption of silicon nitride

The samples studied in our experiments almost always contained a buffer layer of $Si_3N_4$ intermediate between the metallic GdFeCo layer and the substrate. It is therefore important to assess whether optical phonons found in this dielectric layer could contribute to the observed switching of magnetization. The optical phonons of $Si_3N_4$ have been measured extensively by previous groups, with there being a Si-N-Si stretching mode at $\lambda \sim 10.9$ µm [S7] and a Si-N breathing mode at $\lambda \sim 20.8$ µm [S8]. These features are clearly visible in the absorption spectrum of $Si_3N_4$ measured, e.g., by Luke *et al.* in Ref. [S9] (reproduced in Fig. S3). The fact that the absorption peak at $\lambda \sim 10.9$ µm does not coincide with any sign of magnetic switching shown in Fig. 2, either observed in the sapphire- or glass-mounted heterostructure, allows us to exclude the contribution of any phonon modes intrinsic to $Si_3N_4$.

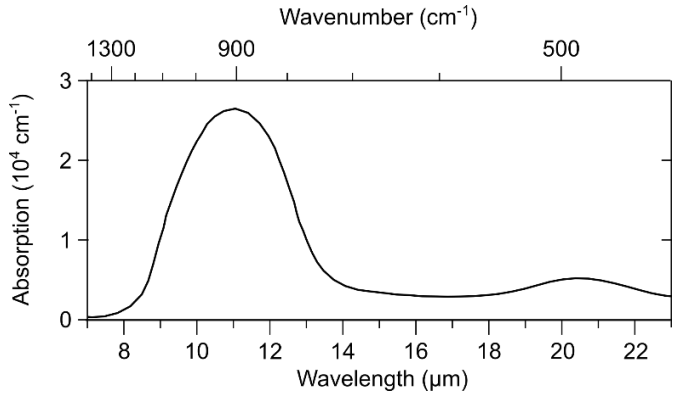

**Fig. S3 |** Spectral dependence of the optical absorption of $Si_3N_4$, calculated using the refractive index obtained by Luke *et al.* in Ref. [S9].

## Supplementary Note 4: Importance of circular / elliptical polarization for magnetic switching

To test the importance of the infrared pulses being circularly-polarized for magnetic switching, we performed additional measurements in which we swept linearly-polarized IR pulses across the sapphire-mounted $GdFeCo/Si_3N_4$ bilayer. The experimental conditions were otherwise the same as those used to obtain the results shown in Fig. 2a. A KRS-5 polarizer was used to ensure linear polarization. The results of these experiments are shown in Fig. S4. Across the spectral range $8 \leq \lambda \leq 21$ µm, we found that the IR pulses only induced demagnetization. This underscores the importance of the impinging excitation being polarized either circularly or elliptically.

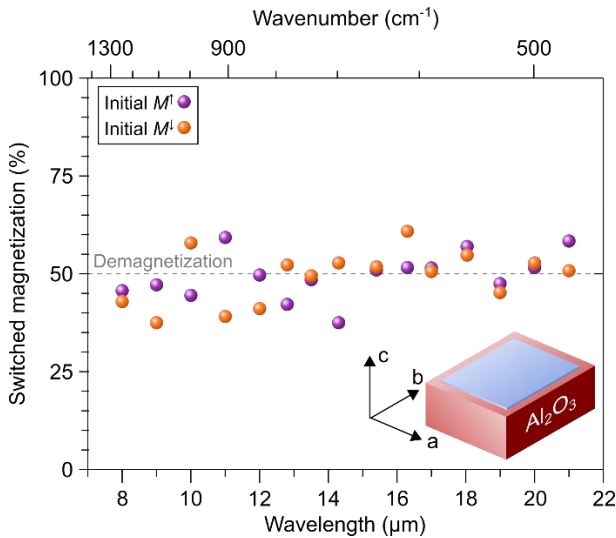

**Fig. S4 |** Spectral dependence of the amount of switched magnetization induced by linearly-polarized IR pulses swept across the sapphire-mounted $GdFeCo/Si_3N_4$ bilayer. The macropulses were scanned across the sample at a speed of 20 µm/s. The purple and orange points correspond to the case where the magnetization was initially pointing up and down respectively.

**Supplementary Note 5: Dependence of the switching efficiency on the sweeping speed**

In Fig. S5, we present the switching efficiency measured as a function of the sweeping speed. We clearly observe that as the speed of sweeping is increased, the efficiency of switching decreases. This is explained by the switching only being achieved at the perimeter of the pumped spot (shown in Fig. 3a). As we therefore move the pumping pulses across the sample at a faster speed, the outer perimeter of the spot has greater opportunity to 'miss' parts of the sample surface.

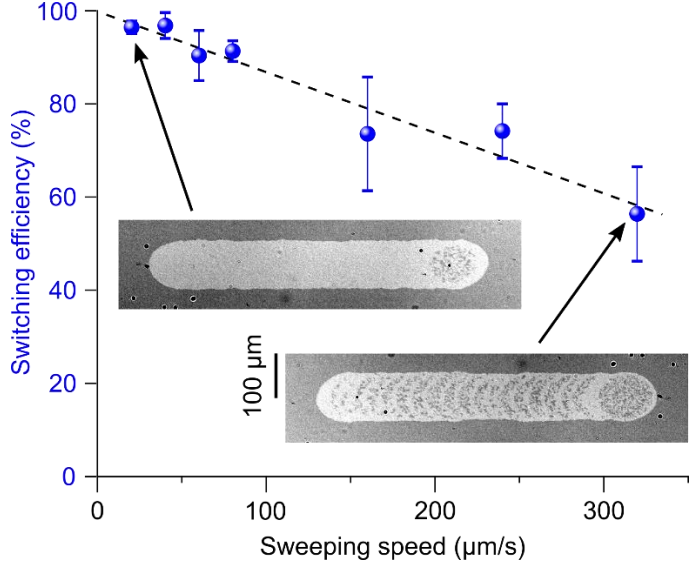

**Fig. S5** | Dependence of the switching efficiency on the sweeping speed of the laser, measured for the sapphire-mounted GdFeCo/$Si_3N_4$ bilayer with $\lambda = 15.5$ µm and duration $\tau \sim 2.0$ ps at room temperature. The insets show typical magneto-optical images recorded with circularly-polarized macropulses with helicity $\sigma^-$.

**Supplementary Note 6: Comparison with other mechanisms of all-optical magnetic switching**

To date, four distinct schemes of all-optical switching of magnetization have been demonstrated. Here, we compare these methods to the discovered wavelength- and helicity-dependent substrate-mediated mechanism of magnetic switching.

(i) Ferrimagnetic alloys of GdFeCo are renowned in the research field of ultrafast magnetism for the possibility to switch its magnetization all-optically using a single laser pulse. This mechanism was first identified by Ostler *et al.* in Ref. [S10], and extensive efforts have since been devoted to understanding its underlying physics. It has now been established that this effect stems primarily from exchange relaxation, which enables angular momentum within the ferrimagnet to be transferred between the antiferromagnetically-coupled sublattices. We refer the reader to Ref. [S11] for an extended discussion. In short, this process is wavelength-independent and helicity-independent. Moreover, it is subject to a cutoff in the duration $\tau$ of the excitation, whereby the magnetization is toggled in polarity if $\tau$ is below a certain threshold ($\tau < \tau_c$), or demagnetized if $\tau > \tau_c$. The precise value of $\tau_c$ depends on the sample composition and the starting temperature.

At room temperature, single-shot experiments with the studied sample of $Gd_{24}(FeCo)_{76}$ mounted on $Al_2O_3$ shows that $\tau_c \sim 1.2$ ps. Since we are able to achieve helicity-dependent all-optical switching of magnetization with pulses substantially longer than $\tau_c$ (see Fig. 1d) only at specific wavelengths, the exchange-driven mechanism of switching cannot explain our results. Moreover, upon raising the starting temperature of our sample to $T_0 = 375$ K, we find that no laser pulse can toggle the magnetization, while the helicity-dependent switching still persists.

(ii) In Ref. [S12], Lambert *et al.* showed that magnetization in certain ferri- and ferro-magnetic systems can be all-optically reversed upon exposure to thousands of circularly-polarized femtosecond pulses with frequencies in the visible spectral range. Subsequent studies have revealed that this comprises a two-step process. First, the magnetization is thermally quenched, leading to the randomly-distributed formation of microscopic magnetic domains. Second, the magnetization is switched by the action of the circular polarization. The physical basis for the latter still represents a subject of intense debate. One argument invokes the inverse Faraday effect (IFE), with the circularly-polarized optical pulse creating a magnetization $\mathbf{M}_{IFE}$ within the magnetic material parallel or antiparallel to its wave vector [S13]. The other argument is based on the material's magnetic circular dichroism (MCD), with the different microscopic domains comprising the multi-domain state absorbing different amounts of energy. Such helicity-dependent temperature gradients can drive domain-wall motion that gives rise to switching [S14].

We exclude this mechanism based on considerations of the IFE and MCD. State-of-the-art *ab initio* models of the IFE predict that the strength of the IFE-induced magnetization increases monotonically with wavelength [S13]. This is entirely contrary to the results observed in Fig. 2, where the switching clearly scales with the absorption spectrum of the underlying substrate. Similarly, MCD is a property intrinsic to the magnetic material of interest, and should be unaffected by the substrate. Indeed, the fact that we do not observe magnetic switching in the silicon-substrate-mounted heterostructure using the same circularly-polarized IR pulses implies that MCD neither drives the switching there nor in the other samples.

(iii) In Ref. [S15], Stupakiewicz *et al.* showed that magnetization can be switched by near-IR optical pulses tuned in frequency to excite specific electronic transitions at resonance, which in turn modifies the crystalline anisotropy. This mechanism relies on the use of linearly-polarized optical pulses. Thus, the fact that the switching shown here relies on the use of mid-IR circularly-polarized optical pulses allows us to exclude this mechanism.

(iv) In Ref. [S16], Stupakiewicz *et al.* showed that mid-IR optical pulses tuned to match the frequency of LO phonon modes can switch magnetization. Since the switching shown in Figs. 1-2 rather relies on the resonant excitation of TO phonon modes, we can exclude the aforementioned mechanism.

**Supplementary Note 7: Characterization of the infrared pump pulses**

The IR pump pulses delivered by the free-electron laser facility FELIX are highly tunable in both wavelength and bandwidth. In Fig. S6a, we present a typical measurement of the laser spectrum recorded during our experiments, overlaid with a Gaussian fit. The standard deviation of the spectrum shown in Fig. S6a is 70 nm, corresponding to a full-width-half-maximum (FWHM) of 166 nm. Since the pulses have been previously shown to be Fourier-transform-limited [S17], we thus estimate the pulse duration to be at least 3.9 ps, corresponding to a minimum of ~56 electric-field cycles.

While the spectrum of the laser is continuously measured, chirp may be introduced by our use of a 5-mm-thick CdSe plate to obtain circular polarization. To therefore assess the impact of the CdSe plate on the duration of the micropulse, we used the technique of sum-frequency generation (SFG) [S18]. An oscillator (MENLO Orange HP), delivering ~200-fs-long pulses with $\lambda = 1040$ nm, is synchronized with the micropulses delivered by FELIX ($\lambda = 9$ μm). The latter is sent across a retroreflector mounted on a motorized translation stage, thus providing temporal resolution. Both lasers are focused onto a 100-μm-thick crystal of $AgGaS_2$. The resulting light produced by SFG, at the central wavelength $\lambda = 932$ nm, is appropriately filtered and measured using a photodiode.

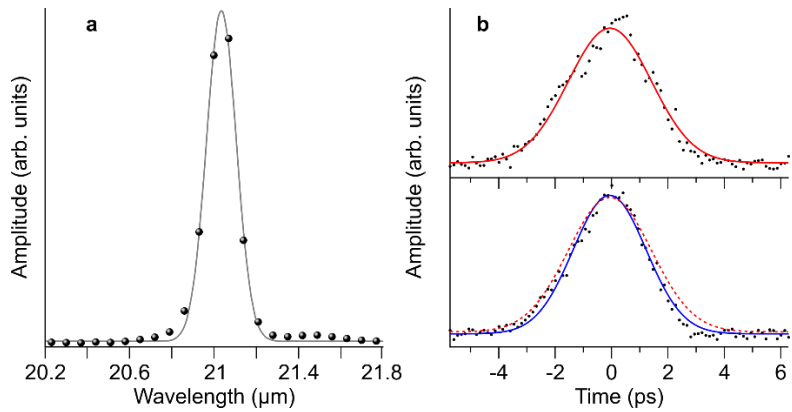

**Fig. S6 | a** Typical spectrum of the IR laser pulse recorded during the measurements. The solid black line corresponds to a Gaussian fit, with a full width at half maximum of 166 nm. **b** Temporal profile of the micropulse ($\lambda = 9$ µm) measured before (red line, top panel) and after (blue line, bottom panel) transmission through the CdSe waveplate. In the bottom panel, the red curve is reproduced to facilitate comparison.

In the top panel of Fig. S6b, we present the SFG signal measured in the absence of the CdSe waveplate. The red line corresponds to a Gaussian fit, yielding a standard deviation of 1.5 ps. Upon insertion of the CdSe waveplate, we observe that the pulse slightly narrows in its duration (bottom panel of Fig. S8b). A Gaussian fit shows the standard deviation is now 1.3 ps. This corresponds to a difference in the spectrum's rms bandwidth of just 5 nm. The CdSe plate is therefore judged as having a negligible effect on the pulse duration, which is in agreement with the known dispersion of CdSe (in the spectral range 8 µm $< \lambda <$ 22 µm, its group velocity dispersion ranges between $-0.2 \times 10^3$ fs$^2$/mm and $-18 \times 10^3$ fs$^2$/mm [S19]). Moreover, as shown in Fig. 3b, we have shown that the pulse duration has no actual effect on the observed helicity-dependent switching of magnetization.

To assess the purity of the circular polarization, we insert a KRS-5 polarizer after the quarter waveplate, and measure the energy of the macropulse using a pyroelectric sensor. In Fig. S7a, we present the normalized energy measured for varying orientations of the KRS-5 polarizer. We clearly observe that the energy of the pulse is broadly independent of the polarizer's orientation, showing that the ellipticity has a ratio in excess of 0.9.

To efficiently drive circularly-polarized optical phonons in the substrate, it is important that the incident circularly-polarized IR light retains its circular polarization upon transmitting through the GdFeCo/$Si_3N_4$ bilayer. This cannot be straightforwardly measured with the sapphire- and glass-ceramic-mounted heterostructures, since immeasurable light is transmitted through the substrates. However, we are able to measure the circular polarization of the IR light transmitted through the silicon-mounted heterostructure. In Fig. S7b, we present the purity of the circular polarization measured before and after transmission through the silicon-mounted heterostructure, obtained in the same fashion as the results shown in Fig. S7a. We clearly observe that the circular polarization is broadly unaffected by the heterostructure, allowing us to conclude that the dielectric/metallic heterostructure does not significantly depolarize the incident light.

The quarter waveplate used to obtain circularly-polarized IR light is nominally functional in the spectral range 1 µm $\leq \lambda \leq$ 22 µm (based on the transmission spectrum of CdSe). To therefore assess the broadband nature of the quarter waveplate, we measured the energy of the infrared pulse for two different helicities as transmitted through the KRS-5 polarizer. In Fig. S7c, we present the measured wavelength-dependent amplitude of the circularly-polarized infrared macropulse. We find that the ratio between these two measurements varies between 0.9 and 1.1, confirming that we are able to obtain circularly-polarized IR pulses across this wide spectral range.

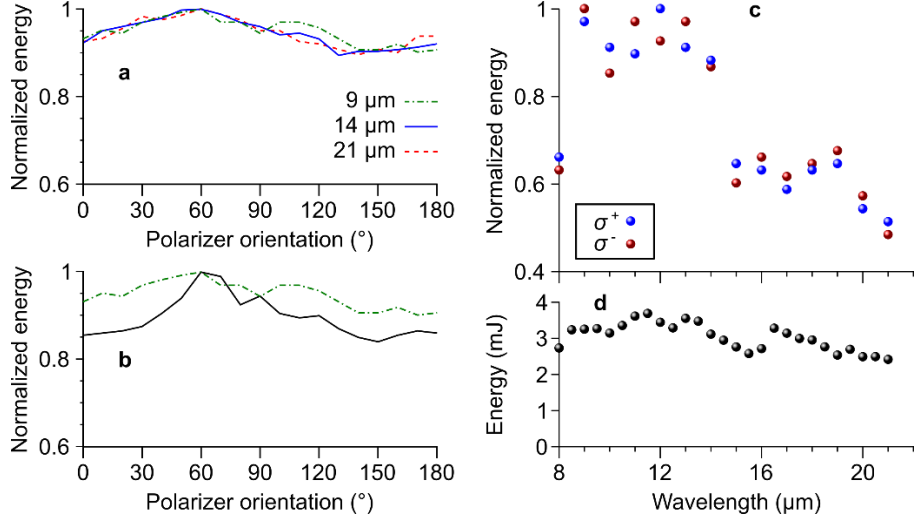

**Fig. S7 | a** Energy of a macropulse after transmission through a quarter-waveplate and KRS-5 polarizer, with the latter rotated to different angles as indicated. The different lines/colors correspond to different central wavelengths as indicated. **b** Similar to panel (a), but with the green dashed-dotted (black solid) line corresponding to the macropulse energy ($\lambda = 9$ µm) before (after) transmission through the silicon-mounted GdFeCo/Si$_3$N$_4$ bilayer. **c** Wavelength-dependent energy of the macropulse with helicity $\sigma^+$ and $\sigma^-$ as indicated, measured after transmission through a KRS-5 polarizer. **d** Typical energy of the macropulse used in the presented results.

Last but not least, while the energy of the IR light delivered by the free-electron laser depends on specific settings of the machine, the energy is broadly independent of wavelength. This is confirmed by the spectral dependence of the laser's energy shown in Fig. 7d.

### Supplementary Note 8: Calculated absorption of infrared light by the heterostructure

In lieu of the optical constants of GdFeCo in the IR spectral range, we have adopted the optical parameters of iron. We justify this assumption by the similar reflectivity dependencies obtained when substituting the GdFeCo layer with an iron layer (see Fig. S2d).

In Fig. S8, we show the results of transfer-matrix calculations performed for the glass-mounted heterostructure [of composition Si$_3$N$_4$(60)/Fe(20)/Si$_3$N$_4$(5)/SiO$_2$ where the number in parentheses is the layer's thickness in nanometers)]. The line in red corresponds to the square of the electric field within the 20-nm-thick layer of iron, revealing an intensity that generally falls with increasing wavelength. This behavior is contrary to the spectrum of switching efficiency observed in the structure (blue dashed line). Furthermore, by fitting the reflectivity of the two heterostructures that are identical other than the substitution of iron with GdFeCo (Fig. S2d), we obtain a scaling function between the two. Upon application of this scaling in the transfer-matrix calculations, we obtain the black line in Fig. S8 characterizing the absorption of IR light by the heterostructure containing GdFeCo. This spectral dependence is again entirely contrary to that of the switching efficiency.

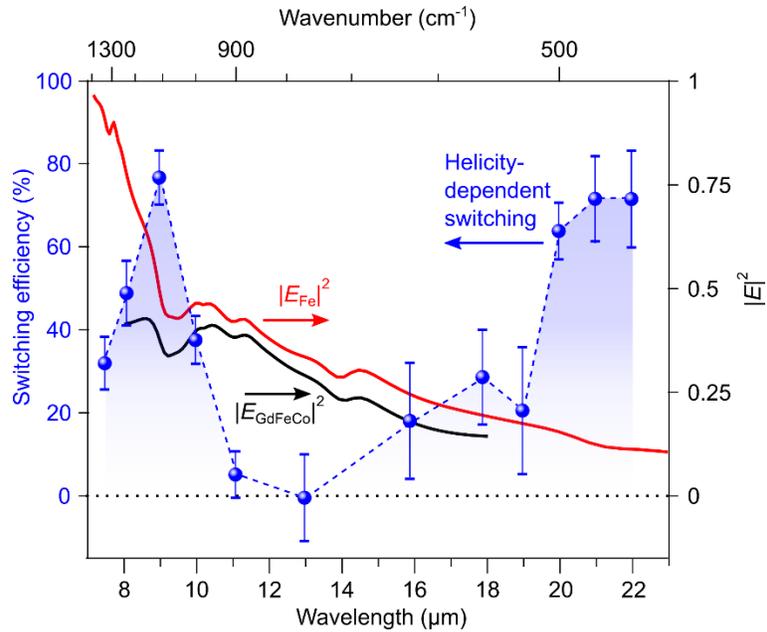

**Fig. S8** | Spectral dependence of the square of the absolute electric field $E$ inside the 20-nm-thick layer of iron (red line) embedded within a heterostucture of composition $Si_3N_4(60)/Fe(20)/Si_3N_4(5)/SiO_2$ (number in parentheses is thickness in nanometers) obtained using transfer-matrix-calculations. The black line is the equivalent spectral dependence when the iron layer is replaced by GdFeCo. To facilitate ease of comparison, the spectral dependence of the helicity-dependent switching of magnetization from Fig. 2b is overlaid.

## Supplementary Note 9: Estimation of the magnetic field generated by the phononic Barnett effect

We must first note that few works have been published on the topic of phonon-induced magnetic fields. Very recent state-of-the-art theoretical works predict field strengths ranging from several milli-Teslas [S20] to several tens of Teslas [S21], although the latter correspond to internal effective magnetic fields stemming from spin-orbit coupling. From our experiments, we can only provide a qualitative estimate of this magnetic field. To do so, we focus the circularly-polarized IR beam on a domain wall in the GdFeCo layer. With a pumping wavelength incapable of driving the substrate's TO phonon mode, no visible effect is found. At resonance, the domain wall moves to one or another side depending on the light's helicity (see Fig. S9). Assuming that the domain-wall motion observed across ~25 µm occurs within 8 µs (the length of the macropulse), we extract a domain-wall velocity on the order of 4 m/s. Upon comparing this with measurements of domain-wall dynamics in the same type of samples (Ref. [S22]), we arrive at magnetic field values on the order of a milli-Tesla.

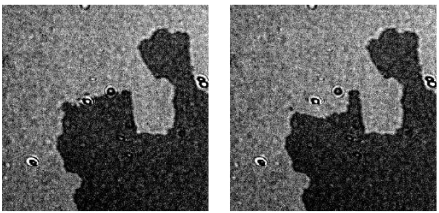

**Fig. S9** | Magneto-optical images taken before (left panel) and after (right panel) irradiation of the sapphire-mounted GdFeCo/$Si_3N_4$ bilayer with a macropulse of wavelength $\lambda = 21$ µm.

**Supplementary Note 10: Importance of managing heat-deposition**

Based on its permittivity, the optical penetration depth of light in sapphire is ~300 nm at $\lambda = 22$ µm. In addition to therefore driving the circularly-polarized TO phonons at resonance, the IR excitation deposits a significant amount of energy converted to heat at the surface of the substrate. This heat, if surviving for a longer timescale than the lifetime of the circularly-polarized phonons, will increase the entropy of the ferrimagnetic layer and destroy its magnetization. Thus, for $\mathbf{M}_{BE}$ to realize the switching, the magnetic nanolayer must be well insulated from the substrate - this is provided by the dielectric $Si_3N_4$ buffer layer.

The above argument is supported by our finding that the helicity-dependent switching is heavily diminished when the $Si_3N_4$ layer is removed (Fig. 3c). In this case, the heat locally generated at the substrate's surface counteracts the impact of $\mathbf{M}_{BE}$. In contrast, with the $Si_3N_4$ buffer layer in place, the ferrimagnetic layer is well isolated from the heating. This helps $\mathbf{M}_{BE}$ to promote the switching.

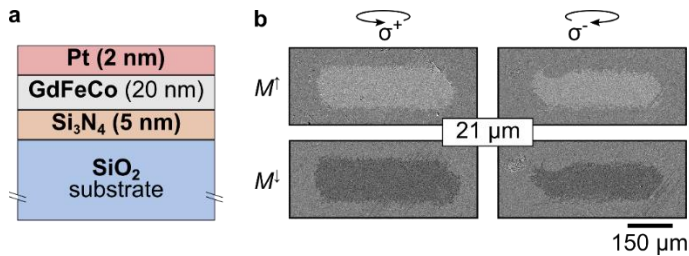

**Fig. S10** | **a** Structure of the sample tested. **b** Background-corrected magneto-optical images taken after sweeping the circularly-polarized macropulses across the sample, with the initial magnetic polarity and optical helicity as indicated.

In addition, we have tested the switching in a glass-mounted GdFeCo/$Si_3N_4$ heterostructure with a 2-nm-thick platinum layer deposited on the top side of the GdFeCo. We have found that the switching previously identified with $\lambda = 21$ µm can no longer be achieved (see Fig. S10). This can be explained by the specific heat capacity of platinum being three times smaller than that of the GdFeCo layer [S23]. As a result, the capping platinum layer is overheated, which in turn overheats the GdFeCo layer for too long. This overcomes the switching driven by $\mathbf{M}_{BE}$.

**Supplementary Note 11: Further elucidation of the inverted helicity-dependent switching**

In Fig. S1b, we highlight in gray the spectral regime in which both the real and imaginary parts of $\varepsilon$ go to zero (~19.5 µm). At this wavelength, we found that the efficiency of helicity-dependent magnetic switching becomes negative in Fig. 2a. This effect is directly observable in Extended Data Fig. 1, when comparing the switching seen in panel (n) to that seen at other pumping wavelengths, e.g., panels (j) and (q). The sense of switching in panel (n) is flipped compared to the sense observed at other wavelengths capable of driving helicity-dependent magnetic switching.

We tentatively explain this result as a consequence of entering the 'epsilon-near-zero' (ENZ) regime [S24]. At the ENZ conditions, 'standard' approaches to linear optics fail while nonlinear interactions reach unity. For example, four-wave mixing processes results in a partial conversion of the beam into a backward-propagating phase-conjugated wave [S25], or in other words, a time-reversed wave [S26]. While the direct wave is spread across the bulk because of its infinite wavelength (i.e., "static optics") [S27], the phase-conjugated one (of opposite helicity) will be dominating at the front surface of the ENZ medium, i.e., exactly at the location where it can influence the magnetic film. This possibly explains why the combination of ENZ and nonlinear optics gives rise to a reversal of the helicity at this excitation frequency.

Further evidence of this scenario comes from the fact that we only observe this inverted sense of switching when focusing the circularly-polarized IR pulses ($\lambda = 19$ µm) exactly at normal incidence on the sapphire-mounted heterostructure. If instead we use an angle of incidence of ~5°, we observe that the switching vanishes. This can be understood as stemming from the light, when impinging on a material under ENZ conditions, undergoing total internal reflection at all finite angles of incidence [S28].

**Supplementary References:**

S1. Barker, Jr, A. S. Infrared Lattice Vibrations and Dielectric Dispersion in Corundum. *Phys. Rev.* **132**, 1474 (1963).

S2. Gervais, F. and Piriou, B. Anharmonicity in several-polar-mode crystals: adjusting phonon self-energy of LO and TO modes in $Al_2O_3$ and $TiO_2$ to fit infrared reflectivity. *J. Phys. C: Solid State Phys.* **7**, 2374 (1974).

S3. Schubert, M., Tiwald, T. E. and Herzinger, C. M. Infrared dielectric anisotropy and phonon modes of sapphire. *Phys. Rev. B* **61**, 8187 (2000).

S4. Nilsson, G. & Nelin, G. Phonon Dispersion Relations in Ge at 80 °K. *Phys. Rev. B* **3**, 364 (1971).

S5. Popova, S., Tolstykh, T. & Vorobev. V. Optical characteristics of amorphous quartz in the 1400–200 $cm^{-1}$ region, *Opt. Spectrosc.* **33**, 444–445 (1972).

S6. Goto, N. et al., U.S. Patent Number 5,391,522, February 21, 1995.

S7. Jhansirani, K., Dubey, R. S., More, M. A. & Singh, S. Deposition of silicon nitride films using chemical vapor deposition for photovoltaic applications. *Results Phys.* **6**, 1059-1063 (2016).

S8. Parsons, G. N., Souk, J. H. & Batey, J. Low hydrogen content stoichiometric silicon nitride films deposited by plasma-enhanced chemical vapor deposition. *J. Appl. Phys.* **70**, 1553 (1991).

S9. Luke, K., Okawachi, Y., Lamont, M. R. E., Gaeta, A. L. & Lipson, M. Broadband mid-infrared frequency comb generation in a $Si_3N_4$ microresonator. *Opt. Lett.* **40**, 4823-4826 (2015).

S10. Ostler, T. A. et al. Ultrafast heating as a sufficient stimulus for magnetization reversal in a ferrimagnet. *Nat. Commun.* **3**, 666 (2012).

S11. Davies, C. S., Mentink, J. H., Kimel, A. V., Rasing, Th. & Kirilyuk, A. Helicity-independent all-optical switching of magnetization in ferrimagnetic alloys. *J. Magn. Magn. Mater.* **563**, 169851 (2022).

S12. Lambert, C.-H. et al. All-optical control of ferromagnetic thin films and nanostructures. *Science* **345**, 1337-1340 (2014).

S13. Berritta, M., Mondal, R., Carva, K. & Oppeneer, P. M. *Ab Initio* Theory of Coherent Laser-Induced Magnetization in Metals. *Phys. Rev. Lett.* **117**, 137203 (2016).

S14. Quessab, Y. et al., Resolving the role of magnetic circular dichroism in multishot helicity-dependent all-optical switching. *Phys. Rev. B* **100**, 024425 (2019).

S15. Stupakiewicz, A., Szerenos, K., Afanasiev, D., Kirilyuk, A. & Kimel, A. V. Ultrafast nonthermal photo-magnetic recording in a transparent medium. *Nature* **542**, 71–74 (2017).

S16. Stupakiewicz, A. et al., Ultrafast phononic switching of magnetization. *Nat. Phys.* **17**, 489–492 (2021).

S17. Knippels, G. M. H. & van der Meer, A. F. G. FEL diagnostics and user control, *Nucl. Instrum. Methods Phys. Res.* **144**, 32-39 (1998).

S18. Knippels, G. M. H. et al., Two-color facility based on a broadly tunable infrared free-electron laser and a subpicosecond-synchronized 10-fs-Ti:sapphire laser. *Opt. Lett.* **23**, 1754-1756 (1998).

S19. Lisitsa, M. P. et al., Dispersion of the Refractive Indices and Birefringence of $CdS_xSe_{1-x}$ Single Crystals. *Phys. Stat. Sol.* **31**, 389 (1969).

S20. Juraschek, D. M., Narang, P. & Spaldin, N. A. Phono-magnetic analogs to opto-magnetic effects. *Phys. Rev. Res.* **2**, 043035 (2020).

S21. Juraschek, D. M., Neuman, T. & Narang, P. Giant effective magnetic fields from optically driven chiral phonons in 4$f$ paramagnets. *Phys. Rev. Res.* **4**, 013129 (2022).

S22. Kim, K. J., Kim, S., Hirata, Y. et al. Fast domain wall motion in the vicinity of the angular momentum compensation temperature of ferrimagnets. Nature Mater **16**, 1187–1192 (2017).

S23. Hellman, F., Abarra, E. N., Shapiro, A. L. & van Dover, R. B., Specific heat of amorphous rare-earth–transition-metal films. *Phys. Rev. B* **58**, 5672 (1998).

S24. Reshef, O., De Leon, I., Alam, M. Z. & Boyd, R. W. Nonlinear optical effects in epsilon-near-zero media. *Nat. Rev. Mater.* **4**, 535–551 (2019).

S25. Pendry, J. Time Reversal and Negative Refraction. *Science* **322**, 71-73 (2008).

S26. Vezzoli, S. et al., Optical Time Reversal from Time-Dependent Epsilon-Near-Zero Media. *Phys. Rev. Lett.* **120**, 043902 (2018).

S27. Kinsey, N. et al. Near-zero-index materials for photonics. *Nat. Rev. Mater.* **4**, 742–760 (2019).

S28. Forati, E., Hanson, G. W. & Sievenpiper, D. F. An Epsilon-Near-Zero Total-Internal-Reflection Metamaterial Antenna. *IEEE Trans. Antennas Propag.* **63**, 1909-1916 (2015).